\documentclass[trackchanges, twocolumn]{aastex701}
\usepackage{amsmath}
\usepackage[thinc]{esdiff}
\usepackage{bigints}
\usepackage{tabularray}
\usepackage{xcolor}
\usepackage{tikz}
\usepackage{graphicx}

\newcommand{\dd}[1]{\text{d}#1}

\begin{document}

\title{A cosmological framework for stellar collisions at high redshift in proto-globular clusters, nuclear star clusters, and Little Red Dots}

\author[orcid=0000-0003-2369-2911,sname='Williams']{Claire E. Williams}
\affiliation{Department of Physics and Astronomy, UCLA, Los Angeles, CA 90095}
\affiliation{Mani L. Bhaumik Institute for Theoretical Physics, Department of Physics and Astronomy, UCLA, Los Angeles, CA 90095, USA}
\email[show]{clairewilliams@astro.ucla.edu}  

\author[orcid=0000-0002-9802-9279,sname='Naoz']{Smadar Naoz}
\affiliation{Department of Physics and Astronomy, UCLA, Los Angeles, CA 90095}
\affiliation{Mani L. Bhaumik Institute for Theoretical Physics, Department of Physics and Astronomy, UCLA, Los Angeles, CA 90095, USA}
\email[]{snaoz@astro.ucla.edu}  

\author[0000-0003-0984-4456]{Sanaea C.\ Rose}
\affil{Center for Interdisciplinary Exploration \& Research in Astrophysics (CIERA), Northwestern University, Evanston, IL 60201, USA}
\affil{NSF-Simons AI Institute for the Sky (SkAI), 172 E. Chestnut St., Chicago, IL 60611, USA}
\email[]{sanaea.rose@northwestern.edu}

\author[0000-0001-5817-5944]{Blakesley Burkhart}
\affiliation{Department of Physics and Astronomy, Rutgers, The State University of New Jersey, 136 Frelinghuysen Rd, Piscataway, NJ 08854, USA \\}
\affiliation{Center for Computational Astrophysics, Flatiron Institute, 162 Fifth Avenue, New York, NY 10010, USA \\}
\email[]{bburkhart@flatironinstitute.org}

\author[0000-0001-7925-238X]{Naoki Yoshida}
\affiliation{Department of Physics, The University of Tokyo, 7-3-1 Hongo, Bunkyo, Tokyo 113-0033, Japan}
\affiliation{Kavli Institute for the Physics and Mathematics of the Universe (WPI), UT Institute for Advanced Study, The University of Tokyo, Kashiwa, Chiba 277-8583, Japan}
\affiliation{Research Center for the Early Universe, School of Science, The University of Tokyo, 7-3-1 Hongo, Bunkyo, Tokyo 113-0033, Japan}
\email[]{naoki.yoshida@ipmu.jp}

\author[0000-0002-8859-7790]{Avi Chen}
\affiliation{Department of Physics and Astronomy, Rutgers, The State University of New Jersey, 136 Frelinghuysen Rd, Piscataway, NJ 08854, USA \\}
\email[]{avi.chen832@gmail.com} 

\author[0000-0002-4086-3180]{Kyle Kremer}
\affil{Department of Astronomy \& Astrophysics, University of California, San Diego; La Jolla, CA 92093, USA}
\email[]{clairewilliams@astro.ucla.edu}

\author[0000-0002-4227-7919]{William Lake}
\affil{Department of Physics and Astronomy, Dartmouth College, Hanover, NH 03755, USA \\}
\email[]{William.Lake@dartmouth.edu} 

\author[0000-0003-3816-7028]{Federico Marinacci}
\affiliation{Dipartimento di Fisica e Astronomia “Augusto Righi”, Università di Bologna, via Piero Gobetti 93/2, I-40129 Bologna, Italy\\}
\affiliation{INAF, Osservatorio di Astrofisica e Scienza dello Spazio di Bologna, via Piero Gobetti 93/3, I-40129 Bologna, Italy\\}
\email[]{federico.marinacci2@unibo.it}

\author[0000-0001-5944-291X]{Shyam H. Menon}
\affiliation{Center for Computational Astrophysics, Flatiron Institute, 162 Fifth Avenue, New York, NY 10010, USA \\}
\affiliation{Department of Physics and Astronomy, Rutgers, The State University of New Jersey, 136 Frelinghuysen Rd, Piscataway, NJ 08854, USA \\}
\email[]{smenon@flatironinstitute.org}

\author[0000-0001-8593-7692]{Mark Vogelsberger}
\affil{Department of Physics and Kavli Institute for Astrophysics and Space Research, Massachusetts Institute of Technology, Cambridge, MA 02139, USA}
\affil{Fachbereich Physik, Philipps Universit\"at Marburg, D-35032 Marburg, Germany}
\email[]{mvogelsb@mit.edu}

\begin{abstract}

Observations and cosmological simulations indicate that the early Universe hosted numerous compact, high-density stellar systems, where close encounters and physical collisions between stars were likely common.
We develop a bottom-up framework for stellar dynamics in such environments, spanning systems with and without intermediate-mass and supermassive black holes, and covering regimes where stellar collisions may or may not dominate the evolution. 
This radially-resolved analytic model connects dense star clusters in their cosmological context to observable outcomes mediated by stellar collisions. 
Initial conditions and environmental properties are drawn from high-resolution cosmological simulations, enabling exploration across a broad region of parameter space.
The analytic predictions are validated against Monte Carlo simulations, demonstrating good agreement across key regimes.
We find that stellar collisions are ubiquitous in many high-redshift environments, with runaway sequences naturally leading to the formation of very massive stars at early times. 
Finally, we show that high rates of destructive collisions can rapidly build up extremely dense gaseous environments around massive black holes, potentially providing an analogue to the observed population of Little Red Dots.


\end{abstract}


\section{Introduction}
Compact environments host a range of important dynamical phenomena, including stellar collisions. 
Locally, these environments include nuclear star clusters, such as the dense cluster surrounding the Milky Way's Sgr A$^*$ \citep[e.g.,][]{schodel_nuclear_2009, genzel_galactic_2010,schodel_surface_2014,feldmeier-krause_triaxial_2017}. 
There, collisions can lead to unusual outcomes, such as the apparent regeneration of an old population with merger products that appear as young massive stars \citep[e.g.,][]{rose_stellar_2023,balberg_stellar_2023,balberg_segregation_2024}. 
The fact that nearly every galaxy hosts a supermassive black hole \citep[see e.g.,][and references therein]{laurikainen_galaxy_2016} suggests that collision-rich environments exist across the Universe. 
As massive black holes and their host star clusters formed, collisions shaped their dynamical and physical evolution.  

The James Webb Space Telescope has opened a window to the era of early galaxy formation ($z\gtrsim 5$), and a wealth of observations that probe systems like and unlike modern galaxies.
Among these observations are objects which demonstrate compactness at various scales and masses at early times. 
An important example are the “Little Red Dots” (LRDs), whose compact size and other observational characteristics suggest that they may host the enshrouded progenitors of supermassive black holes \citep[e.g.,][]{harikane_jwstnirspec_2023, kocevski_hidden_2023,matthee_little_2024,akins_2025_cosmosweb}.
However, not every dense environment at high redshift shows evidence of a black hole.
For example, cluster-scale observations achieved through gravitational lensing show that early star formation was extremely clustered, with massive dense star clusters of surface density upwards of $10^5 M_\odot \, \mathrm{pc}^{-2}$ \citep{Adamo+24,Mowla+24, claeyssens_star_2023,claeyssens_tracing_2025, messa_anatomy_2025,fujimoto_glimpse_2025, fujimoto_primordial_2024, Vanzella+23, whitaker_discovery_2025, Vanzella+22}. 
It may be that these clusters are the high-redshift progenitor systems of globular clusters or nuclear clusters in the local Universe \citep[e.g.,][]{whitaker_discovery_2025,Adamo+24}.  
These systems add a detailed, small scale view to the census of dense galaxies at high redshift \citep[e.g.,][]{Casey+24}. 

From a theoretical perspective, models and simulations suggest that the early Universe naturally produces high density regions. 
For instance, in the most extreme case, compact star clusters may form in a ``feedback-free" burst throughout a high-redshift galaxy, resulting in numerous black hole seeds which can migrate to the central regions \citep[e.g.,][]{dekel_growth_2024, dekel_feedback-free_2025}.
More generally, high halo concentrations—natural at high redshift in $\Lambda$CDM—can lead to efficient star formation and dense systems forming \citep[e.g.,][]{boylan-kolchin_accelerated_2025}.
High resolution cosmological simulations confirm this perspective, finding that even in small clusters and low mass galaxies, extremely high density regions form naturally 
\citep[e.g.,][]{nakazato_merger-driven_2024, Williams+25, williams_observing_2026}.
Simulations find that for the earliest star forming regions at extremely low metallically, when surface densities are high, feedback is strongly counteracted by the extreme gravitational potential \citep[e.g.,][]{grudic_when_2018,2021ApJ...922L...3L, menon_outflows_2023, 2024ApJ...967L..28M, nebrin_lyman-_2025}.
Indeed, theoretical works modeling dense stellar systems find frequent stellar collisions and mergers occurring in early star clusters \citep[or dense systems analogous to these early star clusters, e.g.,][]{devecchi_formation_2009,katz_seeding_2015,Sakurai17, gieles_concurrent_2018,boekholt_formation_2018,vink_very_2023,charbonnel_n-enhancement_2023,reinoso_formation_2023,fujii_simulations_2024,gonzalez_prieto_intermediate-mass_2024, vergara_efficient_2025, vergara_rapid_2025, rantala_frost-clusters_2026, schleicher_massive_2026}. 

For the enigmatic LRD systems, both stellar and AGN models have been proposed to explain the observational features. 
An LRD composed entirely of stars may reach such extreme densities as to explain the observed broad line emission \citep{baggen_small_2024,chisholm_little_2026}. 
Several theoretical works explored the dynamical implications of the stellar interpretation in particular, showing that even if a black hole is not initially present, an LRD composed of a star cluster should result in a dynamically unstable scenario where collapse to a black hole occurs rapidly \citep[e.g.,][]{pacucci_little_2025, vergara_rapid_2025,vergara_efficient_2025, escala_fate_2025, guia_sizes_2024, schleicher_massive_2026}. 
This is consistent with works such as \cite{kritos_supermassive_2024}, which link nuclear star cluster sites in cosmological simulations to the seeds of supermassive black holes via mergers. 
Once a black hole is present, treating an LRD as an early nuclear star cluster analog allows for black hole growth and several interesting dynamical effects, such as Tidal Disruption Events (TDEs) and Extreme Mass Ratio Inspirals (EMRIs) \citep[e.g.,][]{Sakurai19,kritos_nuclear_2025, keitaanranta_rapid_2025, bellovary_little_2025, liempi_constraints_2026}.
Thus, the theoretical literature strongly suggests that LRDs may be some of the most dynamically rich environments in the early Universe, whether or not they host a black hole. 
Recent studies have suggested that LRD observations may be attributed to an actively accreting black hole embedded in extremely dense gas on nuclear scales \citep[e.g.,][]{naidu_black_2025, maiolino_jwst_2025, kido_black_2025,rusakov_little_2026, de_graaff_little_2025}.

Here, we seek to connect LRDs to the broader range of systems that might reach high density at high redshift through a simple and radially-resolved analytical framework. 
In this framework, we explore budding nuclear star clusters, hosting an early massive black hole, and extremely dense cluster/ proto-galaxy cores within the same analytic model. 
We consider a broad parameter space that encompasses proto-globular and nuclear star clusters and the breadth of varying density systems seen in cosmological simulations. 
Thus, with a wide range of initial conditions that do not depend on an observational measurement, we can analytically explore many potential outcomes. 
Our central question is the role of stellar collisions at high redshift--where in the early cosmos did the collisions of stars lead to black hole formation, unusual metal abundances, and other effects? 

Our study offers a novel framework to understand the cosmological environment and prevalence of various system types in a range of parameter space.
This parameter space varies by multiple orders of magnitude in radius, velocity dispersion, and mass, and allows for systems in various stages of dynamical equilibrium.
We include the radial dependence of velocity and density on various density structures, as well as the presence or not of a black hole.
We couple systems' dynamical evolution to their interaction with neighboring halos through comparison to a high resolution cosmological simulation, and check our analytic framework numerically with a Monte-Carlo code.
We find that stellar collisions do occur on the main sequence for a range of systems, analogous to proto-globular clusters or LRDs, leading to enriched gas in the central regions. 
In systems experiencing constructive collisions, we estimate the maximum very massive star (VMS) mass that forms through this process. 
In \S~\ref{sec:analyticmodel}, we describe our analytic approach to the question. 
In \S~\ref{sec:montecarlo}, we describe a set of numerical Monte Carlo simulations that are run to validate this analytic model. 
Our results are presented in \S~\ref{sec:results}. 
A discussion and comparison to the theoretical and observational literature is presented in \S~\ref{sec:comparisontheoryandobs}. 
We summarize the main conclusions in \S~\ref{sec:summary}. 
We use a flat $\Lambda$CDM cosmological model, with $H_0=71$ km/s/Mpc, $\Omega_m=0.27$,  $\Omega_b = 0.044$, $T_{\rm CMB}=2.726$ K and , $\sigma_8 = 0.8$. 

\section{Analytic Framework}
\label{sec:analyticmodel}

\begin{figure}
    \centering
    \includegraphics[trim={5cm 15cm 0 10cm},width=.95\linewidth]{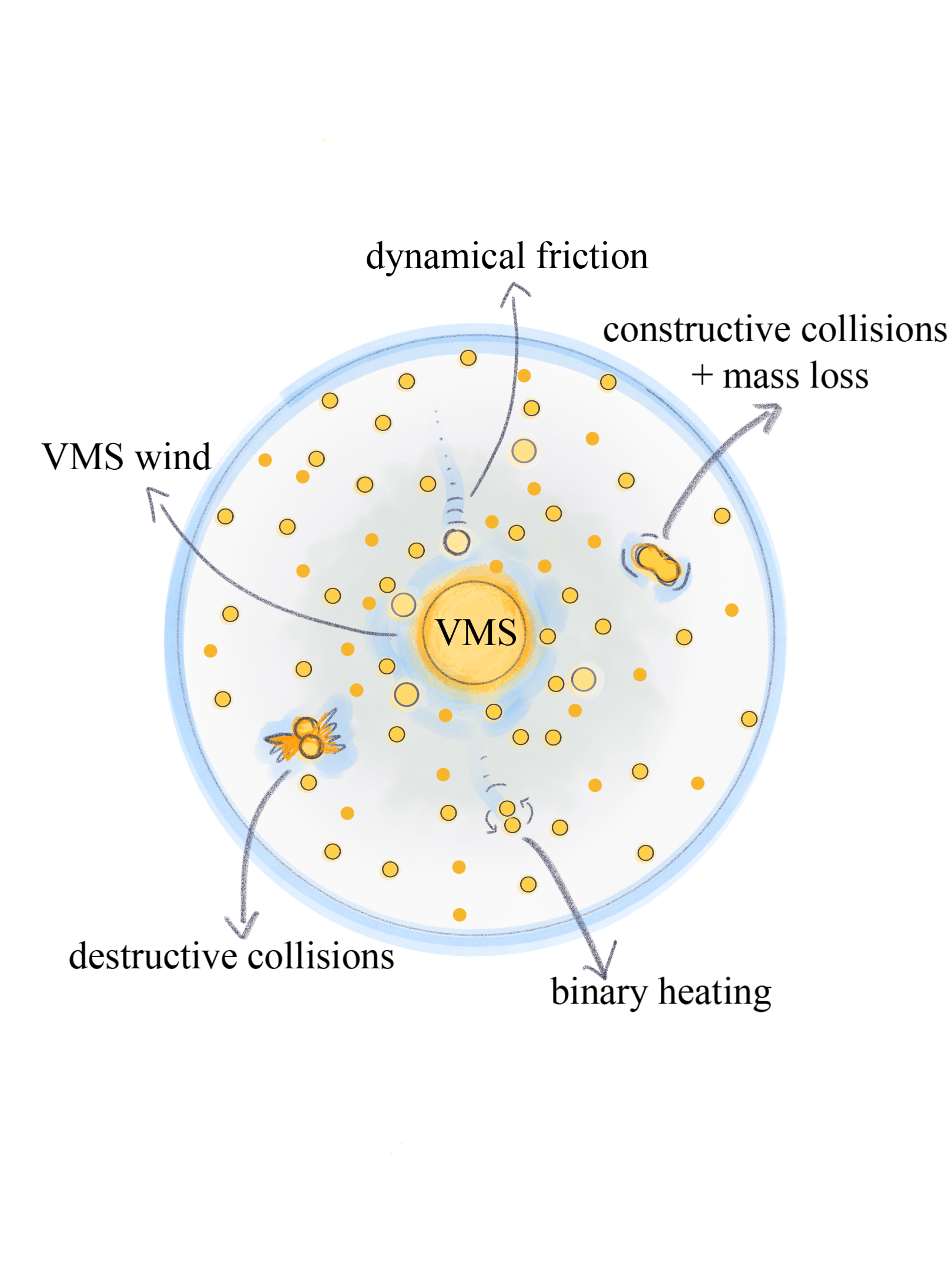}
    \includegraphics[trim={5cm 0cm 0 0cm},width = 0.95\linewidth]{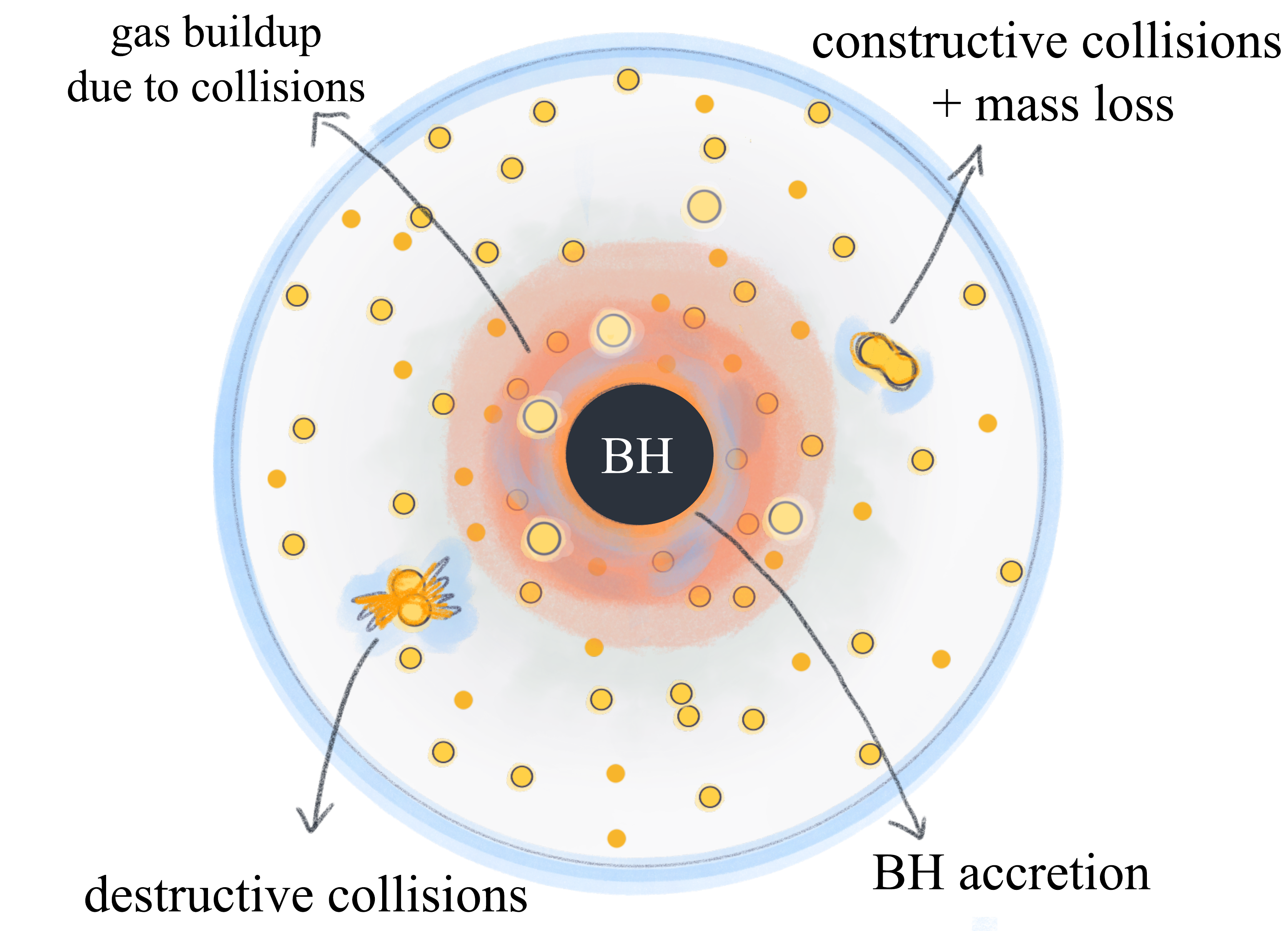}
    \caption{Cartoon schematic diagram of physical processes represented by our analytic model of (top) VMS formation and (bottom) clusters hosting black holes. In both models, we consider the regions of the star cluster that may host destructive versus constructive stellar collisions. For systems without a black hole (top), systems are typically constructive throughout, with massive merger products migrating to the center and contributing to the formation of a VMS. The inward migration is balanced by the effects of depletion and binary heating, as well as winds from the VMS. For systems with a black hole (bottom), we find that the inner regions are dominated by destructive collisions, leading to the buildup of dense gas, which may then accrete onto the black hole or be expelled from the cluster due to feedback, although we do not explicitly model these processes. }
    \label{fig:schematic}
\end{figure}

\begin{figure*}
    \centering
    \includegraphics[width=0.99\linewidth]{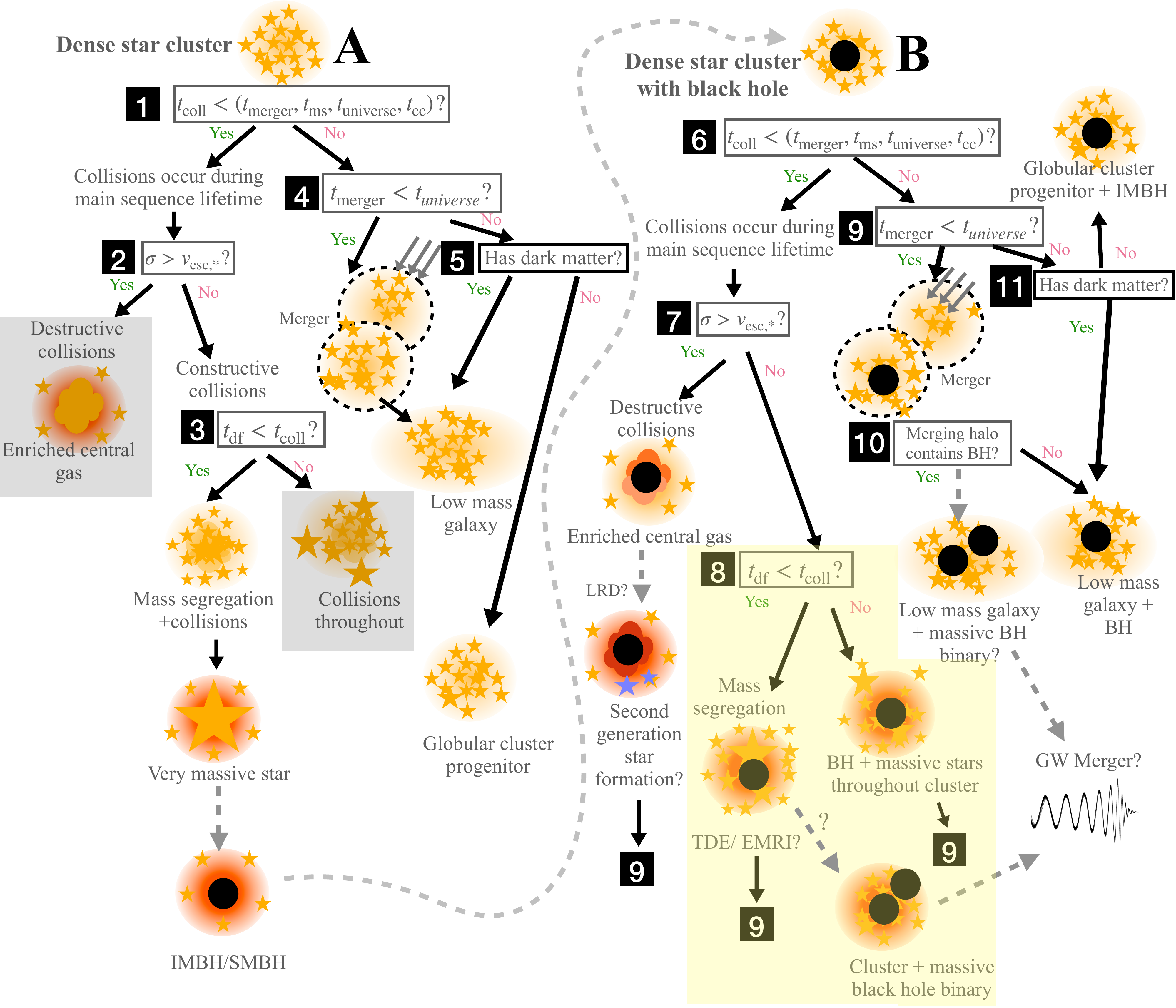}
    \caption{Flowchart depicting the logic of the analytic model considered here. We  consider two initial conditions on the left and right sides: a dense star cluster (left) and a dense star cluster that hosts a massive black hole (right). We are agnostic about the origin of the black hole, although some scenarios on the left hand side can lead to this initial condition.
    Tentative connections such as this are marked with grey dashed arrows. While the flowchart depicts binary either-or decisions, in our analytic calculation, we are able to track mixed outcomes--i.e., radial dependence of the criteria leading to a combination of the scenarios listed. The outcomes marked in grey (at choice (2) and (3)) do not occur for the parameters we consider here. The outcomes highlighted in yellow (choice (8) and below) only occur in systems that also experience destructive collisions in the interior (i.e., yes to choice (7)) and we do not explicitly consider these combined processes in this work.}
    \label{fig:flowchart}
\end{figure*}

In the following sections we describe the analytic framework used to investigate stellar collisions and their outcomes in high redshift systems. 
\autoref{fig:schematic} shows a cartoon diagram of the processes considered for systems with and without a massive black hole. 
Treating each set of initial conditions, we consider these possible dynamical effects through the framework depicted in \autoref{fig:flowchart}. 
In \S~\ref{sec:modelparams}, we describe the initial parameters considered in this work. 
In \S~\ref{sec:disruptionprocess}, we investigate the physical processes that terminate our collision model, either by ending the duration of stellar collisions or disrupting the system significantly away from our initial parameters. 
The dynamical processes depicted in \autoref{fig:schematic} are developed in \S~\ref{sec:collisions}-\ref{sec:heating}. 
Then, we explain the treatment of very massive star formation (\S~\ref{sec:VMS_formation}) and systems with black holes (\S~\ref{sec:model_bh}). 

\subsection{Model parameters}
\label{sec:modelparams}
We investigate a range of systems inspired by the parameters of star clusters found in the cosmological simulations of \cite{Williams+25}, as well as galaxies and star clusters observed at high redshift by {\it JWST}. 
We choose a range in maximum radius spanning $0.2-10^2$ pc, velocity dispersion from $2-200$ km s$^{-1}$ and, mass between $10^4-10^9M_\odot$. 
Clusters may have a black hole ranging from intermediate to supermassive in size. 
\autoref{tab:parameters} summarizes these choices. 
We eliminate any combination of these parameters that leads to a dynamically unstable system or to an average density higher than $6\times10^7 M_\odot$ pc$^{-3}$. 
From this set of parameters, we can imagine modeling the central region of any theoretical or observed high redshift object to be represented by a combination of a star cluster and a black hole. 
This could include a proto-globular cluster, or an LRD hosting a supermassive black hole and a dense star cluster in the inner regions--analogous to our Milky Way's nuclear star cluster. 
\begin{table*}
\centering
\begin{tblr}{
  cell{2}{1} = {r=2}{},
  vlines,
  hline{1-2,4-9,11,12,13,14,15} = {-}{},
}
\textbf{Parameter}                         & \textbf{Comment}                & \textbf{Values Considered}             \\ \hline
\textbf{Typical stellar mass }             & ``Salpeter"      ($\dd N/\dd M\propto M^{-\beta}$)                &        $\beta=-2.35$, $M_*=1M_\odot$                                \\
                                  & ``Top Heavy" ($\dd N/\dd M\propto M^{-\beta}$)                    &   $\beta=-1$ (Log flat), $M_*=17 M_\odot$                                     \\
\textbf{Stellar Density Profile}  & Power Law ($\rho(r)\propto r^{-\alpha}$)~                    & $1.0\leq\alpha \leq 2.0$              \\
\textbf{Black Hole}               & SMBH/IMBH                     & $M_{BH}= 0,1e3,1e4,1e5,1e6,1e7M_\odot$ \\
\textbf{Cluster Radius}           &                               & $0.2 \text{ pc}\leq r \leq 100\text{  pc}$                     \\
\textbf{Cluster Mass}             &                               & $10^4-10^9M_\odot$                                        \\
\textbf{Cluster Velocity}         &                               & $2 \text{ km/s}\leq\sigma\leq 200 \text{ km/s}$        \\
\textbf{Density of neighboring halos} & Nearest neighbor  (``Strict") & ~Sims of \cite{Williams+25}~\\
                                & 10 ckpc mean (``Relaxed") &     \\
\textbf{Final Redshift}         &                               & $12\geq z\geq 5$        \\
\textbf{Stellar Binary Fraction}         &  default $\phi_{\rm bin}=0.2$                             & $0.05<\phi_{\rm bin}<0.2$        \\
\textbf{Stellar Eccentricity}         &  default  $e=0.5$                             & $0<e<0.99$
\end{tblr}
\caption{Table of parameters used in the analytic model in this work. Combinations that lead to an unbound system or average density greater than $6\times10^7 M_\odot$ pc$^{-3}$ are excluded, but systems are allowed to be out of virial equilibrium. For the typical stellar mass, the $M_*$ listed in the third column is used for the stellar mass in the analytic model, while the mass function listed in the middle column is used to calculate the IMF fraction. We test the full IMF in our Monte Carlo Simulations.}
\label{tab:parameters}
\end{table*}
\subsection{Disruption processes}
\label{sec:disruptionprocess}
Given these initial assumptions, we explore the frequency of stellar collisions in various systems. However, various processes (such as the stars in question evolving off the main sequence) may halt the time frame for collisions, and other processes (such as a merger with a nearby halo or the restructuring of the density profile through relaxation) may move the system away from the static initial conditions assumed here. 
Although not all ``disruption" processes will stop collisions, they signify the end of our model's validity, thus we terminate systems at the disruption time. 
In the cases of merger and core collapse, this may lead to an underestimation in the total number of collisions. 
Below we describe the various disrupting effects and our nominal disruption time, which explicitly includes dependence of these processes on the mass and radius of the star cluster systems. 
\paragraph {Cluster-halo merger with nearby objects}
At high redshift, mergers are frequent as hierarchical structure formation proceeds. 
An encounter with a nearby halo will disrupt the stellar system.
The number density of halos is a strong function of mass; thus, we calculate the timescale for a halo of at least 10\% of the mass of the system to come inside the maximum radius of the system. 
 We use the framework of \cite{Williams+25}, which uses typical halo number densities and velocities drawn from a high resolution hydrodynamical simulation. 
 We start with a simple rate calculation: 
 \begin{equation}
     t_{\rm merger} \sim \frac{1}{n_{\rm DM}\sigma_{\rm int} v_{\rm DM}} \ .
     \label{eq:tmerger}
 \end{equation}
 Here, $v_{\rm DM}$, the typical relative velocity between the star cluster and the halo, is taken as $16.6$km s$^{-1}$ at $z=12$, the velocity dispersion of halos in \cite{Williams+25}. 
 The cross section $\sigma_{\rm int}$ is 
 \begin{equation}
     \sigma_{\rm int} = \pi(r_{\rm star cluster} + r_{\rm DM})^2 \ ,
 \end{equation}
 where $r_{\rm star cluster}$ is the star cluster radius $r_{\rm DM}$ is taken to be the virial radius of the dark matter halo:
 \begin{multline}
    r_{\rm vir} = 0.784 \left(\frac{M}{10^8 h^{-1} M_\odot}\right)^{1/3} \left(\frac{\Omega_m \Delta_c}{\Omega_m^z 18\pi^2}\right)^{-1/3} \\ \times \left(\frac{1+z}{10}\right)^{-1}h^{-1} \text{ kpc} \ ,
\end{multline} 
\citep{BarkanaLoeb+01}. 
To find the number density of halos, we use the \citet{ST+02} framework, calibrated to the cosmological simulations' overall number density, $n_{\rm est, sim}$:
\begin{equation}
    n_{\rm DM}(M)= f_{\sigma_8}f_{\rm \delta M}(M)n_{\rm est, sim} \ ,
    \label{eq:numberdens}
\end{equation}
where $f_{\sigma_8}$ is a correction factor for the high value of $\sigma_8$ used in that simulation, $f_{\rm \delta M}$ is the fraction of halos at mass M,
\begin{equation}
    f_{\rm \delta M}(M)=\frac{N(>(M+\delta M))-N(>M)}{N_{\rm tot}} \ .
\end{equation}

\paragraph{Other disruption}
Stars that potentially experience collisions may also evolve off the main sequence prior to the collision timescale.
Thus, we additionally include the main sequence lifetime (as a function of stellar mass) for our systems. 
We also cut off our systems if the core collapse timescale has elapsed $t_{\rm cc}\sim 0.2t_{\rm relax}$ (see \S~\ref{subsec:relax} below), as beyond this time, it is likely that the density profile will restructure and our original profile no longer applies.
Finally, we also include the age of the Universe $t_{\rm Universe}$ at the redshift of interest $z,$ if the halo merger and main sequence timescales are longer. 

Thus, our disruption timescale is:
\begin{equation}
    t_d = \text{min}(t_{\rm merger}, t_{\rm ms}, t_{\rm Universe}, t_{\rm cc}) \ .
    \label{eq:tdisrupt}
\end{equation}
While post merger systems and other disrupted systems may indeed host a range of interesting phenomena, we leave the modeling of their evolution to future work. 

\subsection{Collisions}
\label{sec:collisions}
Stellar collisions may occur on the collision timescale, $t_{\rm coll}.$ 
\begin{equation}
    t_{\rm coll} = \frac{1}{\pi n(r) \sigma(r)}\left( F_1(e) + F_2(e)\frac{1}{\sigma(r)^2} \right)^{-1} \ ,
    \label{eq:t_coll}
\end{equation}
where 
\begin{equation*}
    F_1 = f_1(e)r_c^2 ,
\end{equation*}
\begin{equation}
    F_2 = 2G f_2(e) r_c (M_* + M_c).
\end{equation}
and $f_1(e)$ and $f_2(e)$ are given by \cite{rose_socially_2020}.
This expression includes the effects of gravitational focusing.

\begin{figure*}
    \centering
    \includegraphics[width=0.45\linewidth]{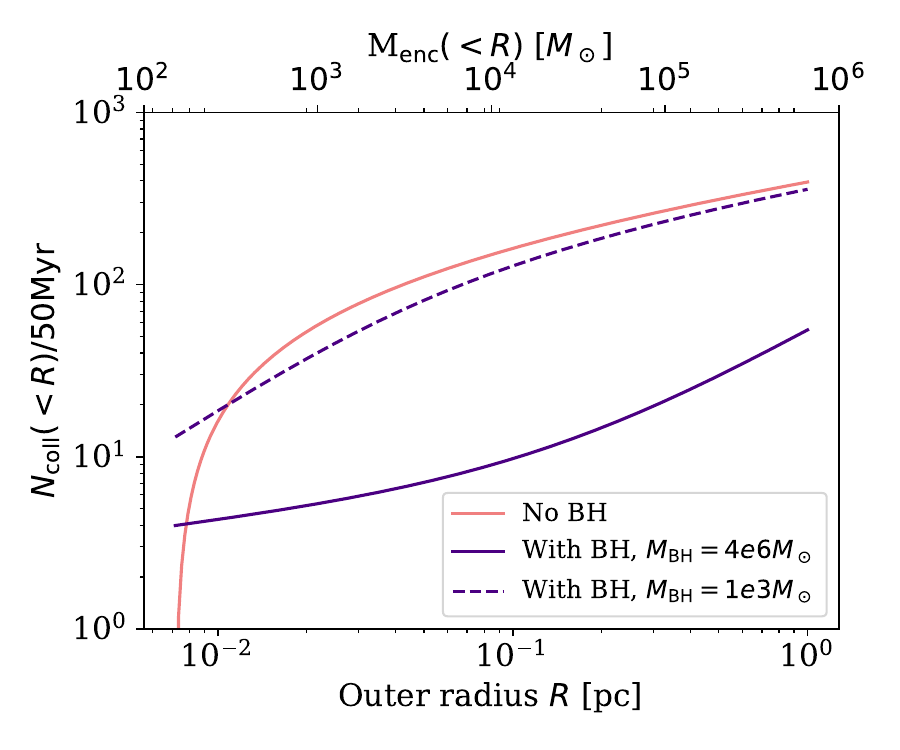}
    \includegraphics[width=0.45\linewidth]{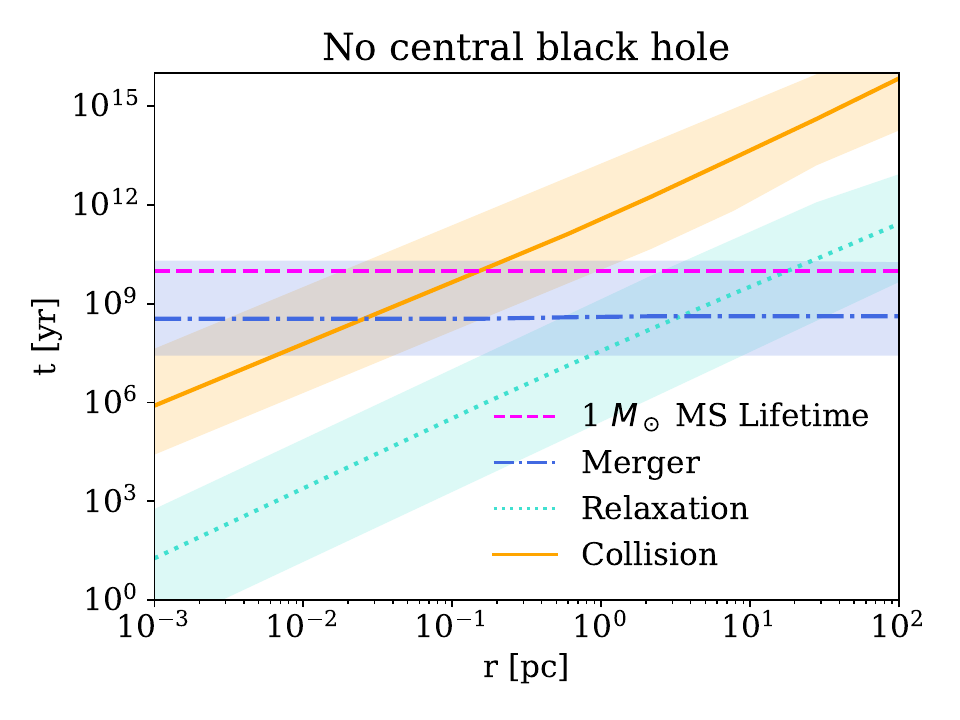}
    \includegraphics[width=0.45\linewidth]{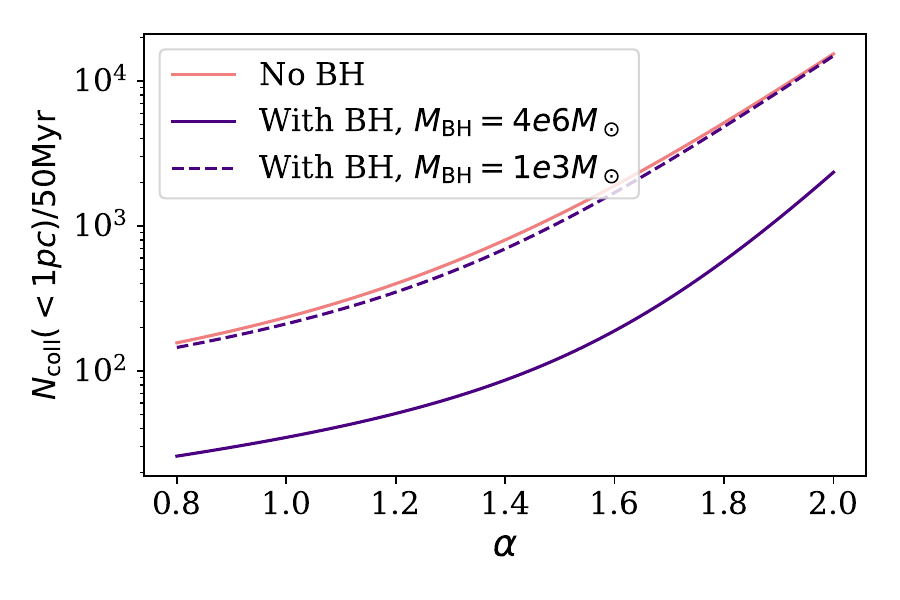}
    \includegraphics[width=0.45\linewidth]{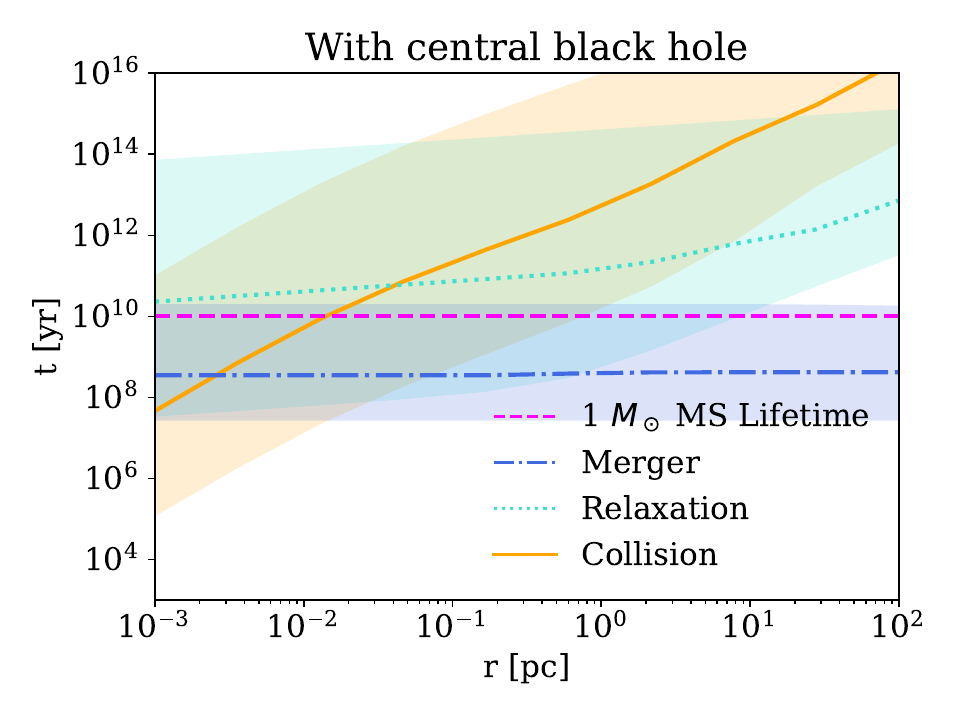}
    \caption{Top left panel: Number of collisions within radius $R$ per 50 Myr, following \autoref{eq:plNumberofCollisions} and \autoref{eq:BHncolltot} for three example cases where the presence and mass of a black hole is varied. In pink, a simple power law with no black hole is shown. In purple, the system has a black hole of $10^3M_\odot$ (dashed) and $10^6M_\odot$ (solid). In all three systems, the power law index $\alpha =1.2$.
    Bottom left panel: Number of collisions within 1 pc per 50 Myr vs power law index, following \autoref{eq:plNumberofCollisions} and \autoref{eq:BHncolltot} for three example cases where the presence and mass of a black hole is varied. In both panels, the cluster parameters are: $\rho_0 = 1e5 M_\odot \text{ pc}^{-3}$, $r_0 = 1$ pc, $f^{\rm IMF}_{M_*}=0.3$, $M_* = M_c =1M_\odot$, $r_{\rm min}=200 R_\odot$, $e=0.5$.
    At discontinuities, the expressions of \autoref{sec:exactsolutions} are used. 
    Right panels: timescale as a function of radius for systems considered in this work without (top) and with (bottom) a supermassive black hole of $10^6M_\odot$. The collision timescale assuming 1 $M_\odot$ stars is solid orange, the main sequence lifetime is pink dashed, the merger timescale (at $z=12$) is dark blue dot-dashed, and the relaxation time is shown in blue dotted. Shaded regions show the range for the systems considered in \autoref{tab:parameters}.}
    \label{fig:exampleNcoll}
\end{figure*}

For a given stellar mass, to find the number of collisions that occur within some radius $r_{\rm max}$, we integrate the number of collisions per star, 
\begin{equation}
    N_{\text{coll}/M_*}(r) = \frac{t_d}{t_{\rm coll}} \ ,
\end{equation}
(using \autoref{eq:t_coll} and \autoref{eq:tdisrupt})
times the number of stars per radius, assuming an IMF where the mass fraction of stars with mass $M_*$ is $f_{M_*}^{\rm IMF}$:
\begin{equation}
    \frac{\dd{N_{M_*}(r)}}{\dd{r}} = \frac{f_{M_*}^{\rm IMF} }{M_*}\frac{\dd{M(r)}}{\dd{r}} \ ,
\end{equation}
over the star cluster, which gives: 
\begin{equation}
    N_{\rm coll} =  \int_{r_{\rm min}}^{r_{\rm max}} \dd{r} N_{\text{coll}/M_*}(r)  \frac{\dd{N_{M_*}(r)}}{\dd{r}} \ .
    \label{eq:GeneralNCollisions}
\end{equation}
The above equation is agnostic to the underlying density distribution--the desired $\rho(r),$ $\sigma(r)$, and $M(r)$ can be inserted. 
We note that the velocity dispersion $\sigma(r)$ reflects stars traveling in an underlying gravitational potential, and thus an expression reflecting the presence of any non-stellar contributions to the potential should be used. 
Meanwhile, the $\rho(r)$ and $M(r)$ reflect the stars alone, so other contributors to the gravitational potential (such as an SMBH) should not be included in those expressions.
For the case of a black hole plus a power law distribution of stars, assuming a steady-state non-rotating system with an isotropic potential and velocity dispersion \citep[e.g.,][]{alexander_distribution_1999},
\begin{equation}
    \sigma^2 (r) = \frac{c_v G}{(1+\alpha)r}(M(r) + M_{\rm BH}) \ .
    \label{eq:sigma_bhpl}
\end{equation}
We integrate this expression analytically and provide the full formulae in Appendix~\ref{sec:exactsolutions}.
For black hole systems, the equilibrium condition results in the \cite{bahcall_star_1976} stellar power law of $\rho\propto r^{-\alpha}$ with $\alpha = 7/4$ \citep[e.g.,][]{linial_stellar_2022, rom_2025_segregation}.
Given that we also consider self-gravitating stellar systems, and systems which may be out of equilibrium, we allow for a range of power law indices between 1-2. Figures are presented for $\alpha=6/5$, a more cored profile than the \cite{bahcall_star_1976} profile, which will result in a more conservative estimate (as demonstrated by the lower left panel of \autoref{fig:exampleNcoll}). We discuss the effect of varying $\alpha$ in the text. 

\autoref{fig:exampleNcoll} shows the behavior of each of these expressions for a characteristic star cluster roughly similar to the Milky Way nuclear star cluster, considering three possible cases regarding the presence of a central black hole. 
The top panel shows the dependence of the number of collisions on radius, and the bottom panel shows the dependence on the power law index. 
This example demonstrates the conflicting effects of the interaction of the black hole with the stellar density profile on the number of collisions.
Because of the gravitational focusing term, a black hole's increased gravitational potential  may suppress or enhance the number of collisions. 


\subsection{Relaxation and mass segregation}
\label{subsec:relax}
The relaxation timescale for a cluster of uniform stellar mass $M_*$ is defined as: 
\begin{equation}
    t_{\rm relax}= \frac{0.34 \sigma^3 }{G^2 \rho M_* \ln(\Lambda)},
    \label{eq:trelax}
\end{equation}
where $\ln(\Lambda),$ the Coulomb logarithm, is given by $\ln(\lambda N)$ \citep{BT08}. 
In cluster dynamics, it is common to consider two regimes for collisions. 
When the collision timescale is shorter than the relaxation timescale, the cluster cannot dynamically respond to the collisions. 
When the relaxation timescale is shorter, stars may diffuse in and out of the regions where collisions are occurring, providing a renewed source of potential colliders of the original mass. 
The relaxation timescale is plotted as a function of radius on the right side of \autoref{fig:exampleNcoll}, for a system with a power-law density profile without (top) and with (bottom) a central supermassive black hole. 

The local dynamical friction timescale for an object of mass $M$, moving within a sea of stars of mass $M_*$, is 
\begin{equation}
    t_{\rm df}\approx \frac{M_*}{M}t_{\rm relax}.
    \label{eq:tdynfriction}
\end{equation}
For a given system, if constructive collisions occur in the outskirts leading to the formation of more massive objects, we check whether the dynamical friction time is shorter than the collision timescale. 
When this is the case, massive collision products may efficiently migrate inward through dynamical friction.

When no black hole is present, the relaxation time becomes short (see \autoref{fig:exampleNcoll}). 
Since we consider secular processes, we set a minimum radius for calculations that include relaxation or dynamical friction by setting the relaxation time equal to the orbital period.

\subsection{Collision outcomes}
\label{sec:collisionoutcomes}
Stellar collisions may be constructive or destructive, depending on the relative velocity at which the stars encounter one another ($\sigma_{\rm rel}$). 
When 
\begin{equation}
    \sigma_{\rm rel}>v_{\rm esc,*},
    \label{eq:destructivecriterion}
\end{equation}
where $v_{\rm esc,*}$ is the escape velocity from the surface of the stars, then the collision will result in significant mass loss--typically destroying the stars involved. 
Such collisions release material of the inner layers of the stars into the local region. 

If instead the collision occurs with a relative velocity lower than the escape velocity, the collision is assumed to be a constructive merger, resulting in the formation of a single, more massive star. 
Following \cite{lai_collisions_1993}, the final mass of the resulting collision is given by $M_f = (1-f_{\rm ml})(M_i + M_*)$, where $f_{\rm ml}$ is the mass loss fraction during the collision, which depends on the energy ratio of the colliding stars (see \cite{rose_stellar_2023} for details): 
\begin{equation}
    f_{\rm ml}=\frac{\mu \sigma^2}{GM_i^2/R_i+GM_*^2/R_*^2},
    \label{eq:masslossfraction}
\end{equation}
where $\mu$ is the reduced mass and $R_i,R_*$ are the stellar radii involved in the collision. 
In the mass loss only scenario, the collision is destructive and the colliding star is not captured. 
In this case, $M_f = (1-f_{\rm ml})M_i$. 
Typically, only the systems with a black hole reach high enough velocity dispersions to achieve significant destructive collisions in the inner regions. 

From here, we may estimate analytically the rate of collision products that rain down on the central region of the cluster due to dynamical friction. 
The rate of mass infall due to dynamical friction ($\dot{M}_{\rm df}$ is modulated by a depletion rate ($\dot{M}_{\rm dep}$), representing the depletion of collision inputs over time:
\begin{equation}
    \dot{M}_{\rm mig} = \dot{M}_{\rm df}- \dot{M}_{\rm dep}.
\end{equation}
Thus, $\dot{M}_{\rm mig}$ represents the overall infall of collision products to the central region
We calculate these expressions using: 
\begin{equation}
    \dot{M}_{\rm df} =\int_{\rm rmin}^{\rm rdf} \frac{\dd{N_{\rm coll}}}{\dd{r}} \frac{M_{\rm coll}(r)}{t_{\rm df}(r)}
    \dd{r},
    \label{eq:df_mig}
\end{equation}
and 
\begin{equation}
    \dot{M}_{\rm depletion} = \int_{\rm rmin}^{\rm rdf} \Gamma_{\rm coll}M_{\rm coll}(r) 4 \pi r^2 n(r)\dd{r}.
    \label{eq:dep_mig}
\end{equation}
Once again, we include the full versions of these expressions in the Appendix (\ref{sec:exactsolutions}). 
\subsection{Binary heating}
\label{sec:heating}
Another important dynamical effect preventing the collapse of the cluster is binary heating.
We formulate the inclusion of this effect as an energy production rate of binaries interior to the dynamical friction radius $\dot{E}_{\rm bin}$.  
The contribution of this effect to the rate of mass inflow will be expressed as
\begin{equation}
    \dot{M}_{\rm bin} \approx \frac{\dot{E}_{\rm bin}}{\Delta \Phi},
    \label{eq:mdotbin}
\end{equation}
where $\Delta \Phi$ is the energy per unit mass required to move matter outward from the center to a few times the dynamical friction radius ($\Delta \Phi\sim \eta \sigma_{r = r_{\rm df}}^2$). 
We use the volumetric binary heating rate from \cite{vesperini_range_1994}:
\begin{equation}
    \epsilon =\frac{G^2 m^3}{\sigma}n^2(\mu_{bs}+\mu_{bb}),
\end{equation}
where $\mu_{bb}$ and $\mu_{bs}$ are quantities representing the average rate of energy release by pairwise interactions between binary-binary and binary-single interactions, respectively. We follow \cite{vesperini_range_1994} to numerically evaluate these factors (their Eqs. 2-3) using the results of \cite{hut_binary-single_1983} and \cite{heggie_binary-single-star_1993}.
This allows us to vary the binary fraction. 
Integrating over volume and inserting into \autoref{eq:mdotbin}, 
\begin{equation}
    \dot{M}_{\rm bin}= \frac{1}{ \eta \sigma_{r = r_{\rm df}}^2}\int_{\rm rmin}^{\rm rdf}  \frac{G^2m^3}{\sigma(r)}n(r)^2(\mu_{bs}+\mu_{bb} )\dd{V}.
    \label{eq:binarymassoutflow}
\end{equation}


\subsection{Very Massive Star Formation}
\label{sec:VMS_formation}
In the cluster center, collisions between these products may result in the formation of a Very Massive Star (VMS). 
We follow a similar framework to \cite{pacucci_little_2025} to find an equilibrium mass for the growing star. 
This equilibrium represents a balance between mass loss due to stellar winds and accretion of new mass: 
\begin{align}
    \frac{\dd{M_{\rm VMS}}}{\dd{t}} &=\dot{M}_{\rm acc}-\dot{M}_{\rm wind}(M,Z),
    \\&  = (1-f_{\rm ml, VMS})(\dot{M}_{\rm df}-\dot{M}_{\rm dep}-\dot{M}_{\rm bin})\nonumber \\&\text{       } -\dot{M}_{\rm wind} \ ,    \nonumber
    \label{eq:fullvmsrate}
\end{align}
where the stellar wind mass loss rate is given by \cite{vink_very_2018}: 
\begin{multline}
    \log{\dot{M}_{\rm wind}}= -9.13+2.1 \log(M/M_\odot)\\+0.74\log(Z/Z_\odot) [M_\odot \text{ yr}^{-1}] \ , 
\end{multline}
and we use the full forms of $\dot{M}_{\rm df}$, $\dot{M}_{\rm dep}$, and $\dot{M}_{\rm bin}$ given in \autoref{eq:fullmassaccretionratepl}, \autoref{eq:fulldepletionrate}, and \autoref{eq:fullbinaryrate}, and have assumed $Z=0.1Z_\odot$. 
The VMS grows due to accretion of stars until an equilibrium is reached at $\frac{\dd{M}}{\dd{t}}=0$. 
The factor (1-$f_{\rm ml, VMS}$) represents the fraction of mass from the inward migration that winds up accreting onto the VMS and any mass lost during that collision. 
We use an energy balance between the energy at the Roche limit of the very massive star and the binding
energy of the infalling star and VMS, similar to \autoref{eq:masslossfraction}.
\begin{equation}
    f_{\rm ml, VMS}=\frac{\mu \sigma^2}{GM_{\rm VMS}^2/R_{\rm VMS}+GM_*^2/R_*},
\end{equation}
For a VMS with radius given by:
\begin{equation}
    R_{\rm VMS}= 2600 R_\odot \left(\frac{M_{\rm VMS}}{100 M_\odot}\right)^{1/2},
\end{equation} 
\citep[][]{hosokawa_formation_2013}, we calculate the velocity at the Roche limit:
\begin{equation}
    \sigma^2 = \frac{2 G M_{\rm VMS}}{R_{\rm Roche}}.
\end{equation}
Typical values of $f_{\rm ml, VMS}$ fall below $10^{-3}$ for a solar mass collider, however  $f_{\rm ml, VMS}$ can reach $10^{-2}$ for a more massive star or when the VMS is below 100$M_\odot$, so we use $2\times 10^{-2}$ as a conservative upper limit. 
We choose to remain conservative regarding this accretion efficiency, given the inherent uncertainty in estimating the structure of stars with such extremely high masses. 
The model is summarized in the schematic of \autoref{fig:schematic}.
Our maximum radius of integration is the dynamical friction radius, $r_{\rm df},$ where $t_{\rm df}<t_{\rm disrupt}$.

This prescription leads to the formation of a range of massive stars that span the typical range of ordinary massive stars (10s of $M_\odot$) to very massive stars to supermassive stars. 
Some systems do not reach $10M_\odot$. 
For simplicity we denote the product a ``very massive star" (VMS) to distinguish from the typical stars in the cluster, but we caution that our model terminates with a wide range of masses.
In this steady-state balance, we have ignored the fact that accretion and inflow onto this star is inherently time dependent, which could lead to episodic or runaway growth rather than this equilibrium limit. 

\subsection{Systems with a massive black hole}
\label{sec:model_bh}
As described above, the presence of a black hole increases the velocity dispersion in the innermost regions, decreasing the effect of gravitational focusing and providing significant orbital energy for destructive collisions. 
We consider the mass loss radius, wherein collisions are destructive, from the criterion given by \autoref{eq:destructivecriterion}:
\begin{equation}
    r_{\rm ml, BH}=
    \frac{c_v G M_{\rm BH}}{v_{\rm esc}^2(1+\alpha)}.
\end{equation}
Evaluating \autoref{eq:GeneralNCollisions} from the tidal radius to $r_{\rm ml, BH}$ gives us the number of destructive collisions that occur before $t_{\rm disrupt}$, and multiplying by $M_*$ gives the mass converted to gas through collisions ($M_{\rm g,tot}$).
Next, we provide a rough estimate for the density of gas in the region around the black hole contributed by collisions by averaging over the volume inside a given radius:
\begin{equation}
    \rho_{\rm g, collisions}\approx
    \frac{4\pi 
    N_{\rm coll, des}(r)M_*}{3 r^3},
\end{equation}
where $N_{\rm coll, des}$ is the number of destructive collisions. We ensure that no more mass is injected than the cluster has available, i.e., we take $\max(M_{\rm g,tot},M_{\rm enc}(r_{\rm ml, BH}))$. 
This approximate value does not reflect the fact that the gas in the region may be highly non-uniform, as it originates through collisions, and we have not included the effects of feedback and cooling. 
Furthermore, supernova feedback may serve to rapidly disrupt the gas. 
Thus, for systems with a black hole, we add an additional disruption time $t_{\rm SN}\sim 50 $ Myr, after which we terminate the additional buildup of gas. 

\section{Monte Carlo Simulation}
\label{sec:montecarlo}
To test our analytical framework above, we simulate a subset of scenarios using a Monte Carlo approach following \cite{rose_stellar_2023, rose_orbital_2025}. 
The aim of these simulations is not to carry out a full N-body validation of the framework for stellar collisions. 
Rather, we seek to test how well we have identified the relevant parameter space within objects where collisions are rampant, and whether or not they are destructive.
In particular, this code does not self-consistently update the masses and number densities of the background stars to reflect the outcome of collisions. 

\subsection{Monte Carlo Method}
We utilize the code developed by \cite{rose_stellar_2023} and \cite{rose_orbital_2025} for systems with a central SMBH and adapt it to a variety of cluster environments with different mass density profiles and velocity dispersions, to understand these systems in their cosmological context. 
We initialize 1000 stars for each test case, located within 1 pc of the center. 
We draw their masses from the IMFs in \autoref{tab:parameters}. 
We draw their initial semi-major axis from a log uniform distribution, to sample inner and outer regions of the parameter space. 
We draw orbital eccentricities from a uniform distribution between $0$ and $1$. 
The simulation includes a statistical approach to collisions, comparing a randomly drawn number to the probability $\Delta t/ t_{\rm coll}$ that a star will experience a collision during timestep $\Delta t$. 
In general, we allow for two outcomes if a collision occurs—merger or mass loss. 
In a merger, the star collides with a star of mass equal to the peak mass or typical mass ($M_*$, also given in \autoref{tab:parameters}) of the chosen IMF.

We evolve our systems for the disruption time (\autoref{eq:tdisrupt}) and use $\Delta t= P_{\rm orbit}$, the Keplerian orbital period.
During each timestep, this code also accounts for the two-body relaxation kicks accrued by applying a small velocity kick, drawn from a Gaussian distribution with $\sigma = \Delta v_{\rm rlx}/\sqrt{3} = v_{\rm orbital}\sqrt{P_{\rm orbital}/t_{\rm rlx}}$ \cite{rose_stellar_2023}. 
The full set of equations is given in \cite{lu_supernovae_2019}. 
The framework of \cite{rose_stellar_2023} approximates the orbits and velocity dispersion as Keplerian using the central black hole mass. In order to investigate the ``no black hole" case in our model above, we use a modified version where the Keplerian orbital central mass is approximated by a small ($20 M_\odot$) black hole plus the enclosed mass within the star's semi-major axis.

\subsection{Analytic model validation}
To test our framework for where destructive collisions occur, we run ten iterations listed in \autoref{tab:montecarlotable}, varying the underlying density profile and black hole mass. 
We confirm that our cutoff radius for destructive versus constructive collisions (\autoref{eq:destructivecriterion}) is consistent with these simulations.
To do this we check whether the analytic $r_{\rm ml}$ agrees with the radius within which $99\%$ of systems lost mass over the course of a collision (rather than gained). 
The typical agreement is within a factor of a few. (see App~\ref{app:montecarlo}).
We also study whether the number of collisions experienced by each star is consistent with our model. 
An example of these tests is also shown see App~\ref{app:montecarlo}. 
Generally, we have good agreement at large radii, whereas the analytic model predicts more collisions in the innermost radii than occur in the simulation. 
The effects of dynamical friction increasing the typical mass of stars in the inner regions are not captured in the simulation, so the typical number of collisions at small radii is larger than what is predicted by the Monte Carlo method. 
With these checks in hand, we continue to investigate the range of outcomes from our analytic model below.

\begin{table}[]
    \centering
\begin{tabular}{|c|c|c|c|c|}
\hline
     & ${M_{\rm BH}}$ ($M_\odot$) & $\rho_0$ ($M_\odot$/ pc$^3$) & $r_0$ (pc) & $\alpha$ \\ \hline \hline
     1 & $4\times 10^6$ & $3 .684 \times 10^3$ & 3 & 1.75 \\ \hline
    2 & $ 10^7$ & $3 .684 \times 10^3$ & 3& 1.75 \\ \hline
    3 & $4\times 10^6$ & $10^5$ & 1& 1.75 \\ \hline
    4 & $ 10^7$ & $10^5$ & 1& 1.75 \\ \hline
    5 & $ 10^5$ & $10^5$ & 1& 1.75 \\ \hline
    6 & $ 10^5$ & $10^5$ & 1& 1.2 \\ \hline
    7 & $4\times 10^6$ & $10^5$ & 1 & 1.2 \\ \hline
    8 & * $2\times 10^1$ & $10^5$ & 1 & 1.75 \\ \hline
\end{tabular}
    \caption{Parameters used in  simulation runs. * indicates systems run with the modified code described in \S~\ref{sec:montecarlo} to approximate a ``no-bh" case.}
    \label{tab:montecarlotable}
\end{table}

\section{Results}
\label{sec:results}
\subsection{System outcomes}
Using the analytic arguments of the previous section, we explore the potential outcomes for a dense star cluster in \autoref{fig:flowchart}. 
We consider two parallel tracks--systems with and without black holes. 
First, the left  case of star clusters with no initial black hole is discussed in \S \ref{sec:runawaycollisions}. 
The right side--scenarios with a black hole--is explored in \S \ref{sec:black holes}. 
The outcome of significant stellar collisions can be the formation of a black hole, so we draw a tentative line in \autoref{fig:flowchart} between the formation of an IMBH and the start of a cluster system containing a black hole.
However, as the processes leading to black hole formation will significantly restructure the density profile of the system, we do not continue the evolution of the non-black hole systems through the black hole track. 
We remain agnostic as to the origin of the black hole in the system.


\subsection{Runaway collisions and Very Massive Star Formation}
\label{sec:runawaycollisions}

Based on the timescales plotted in the upper right panel of \autoref{fig:exampleNcoll}, we posit that collisions should be prevalent in a range of star cluster or nuclear cluster environments at high redshift. 
Following point (2) in our flow chart (\autoref{fig:flowchart}), the next criteria to consider are whether destructive collisions occur in any part of the object, and for constructive collisions, whether the products migrate to the central region due to dynamical friction. 
We find that no systems in the no-black hole regime have deep enough potential wells to allow for destructive collisions. 
Additionally, the dynamical friction time is extremely short, resulting in all systems experiencing mass segregation of collision products in at least the central regions. 
\begin{figure*}
    \centering
    \includegraphics[width=0.46\linewidth]{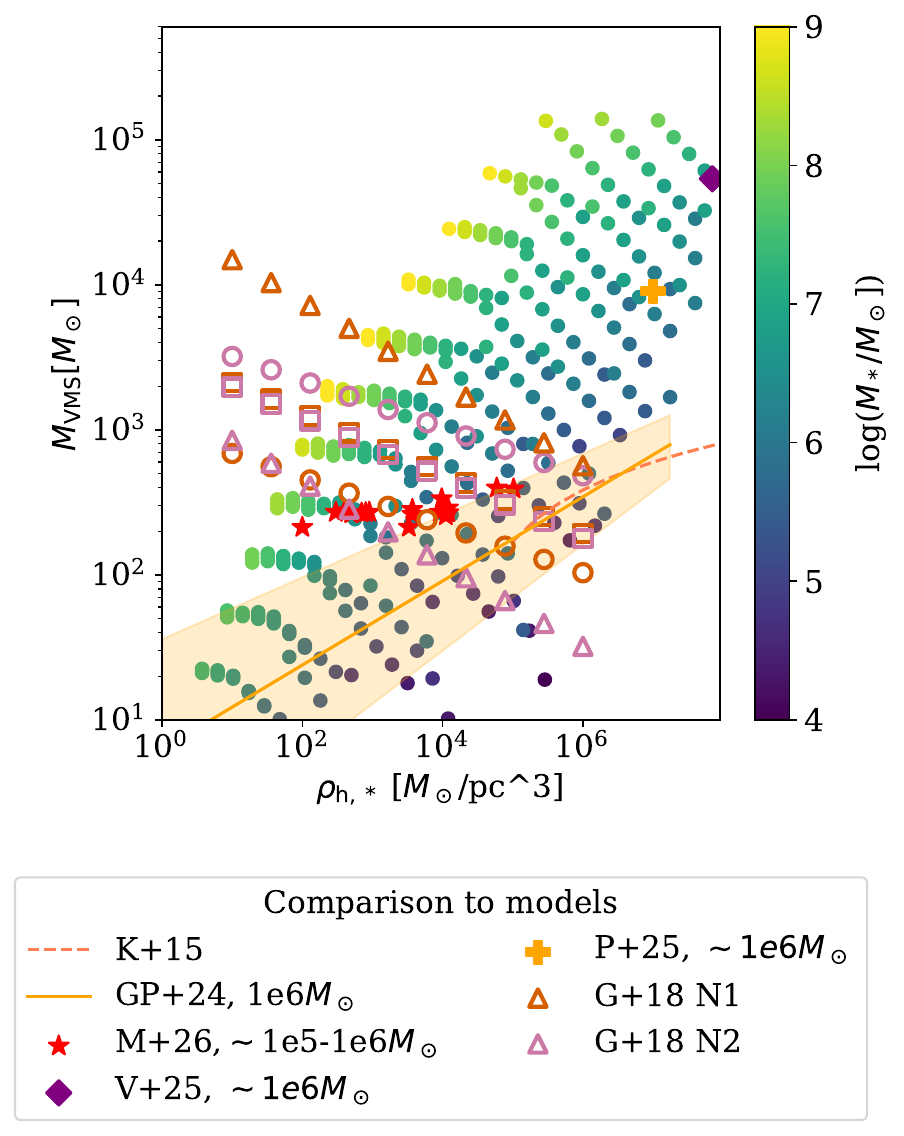}
    \includegraphics[width = 0.49 \linewidth]{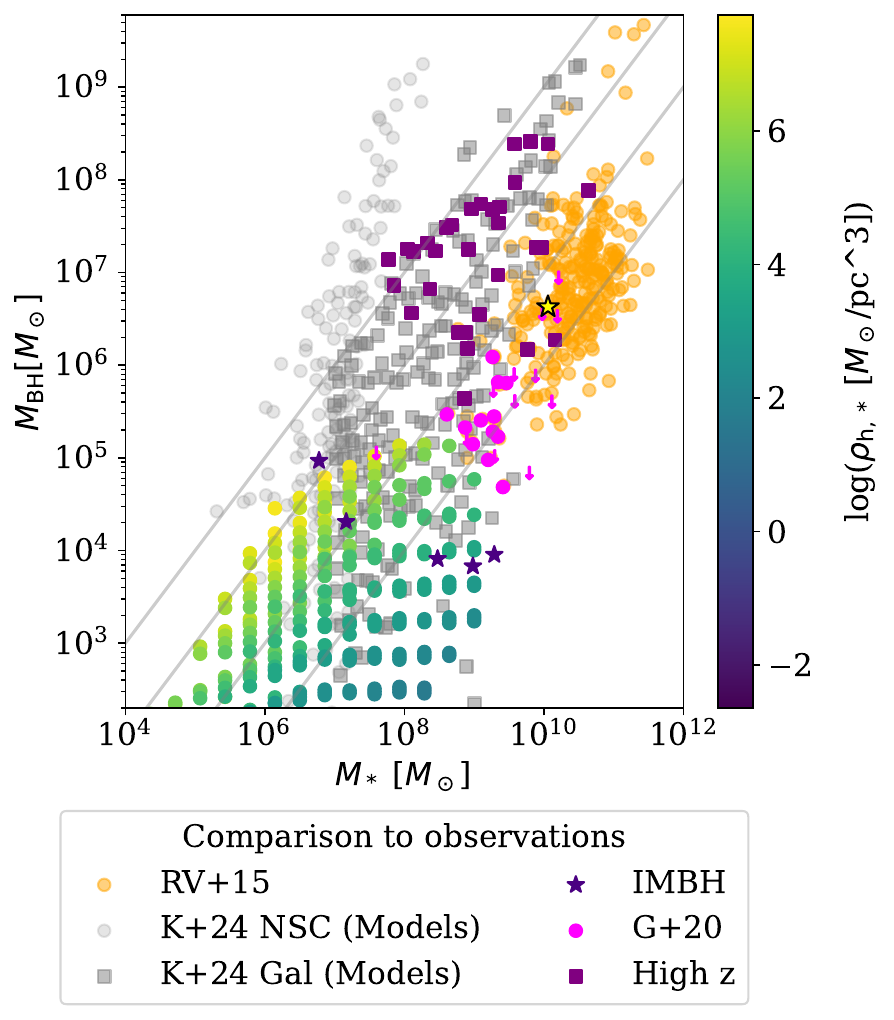}
    \caption{{\bf Left:} Maximum very massive star (VMS) mass versus cluster half-mass density $\rho_h,*$, compared to theoretical predictions. The color bar depicts the total cluster mass. For comparison, we plot the fitting function of \cite{gonzalez_prieto_intermediate-mass_2024} in orange for a $10^6 M_\odot$ star cluster of varying density and the simulations of \cite{mestichelli_teen_2026} as red stars, \cite{vergara_rapid_2025} as the purple diamond, and \cite{pacucci_little_2025} as the orange plus, the simulations of \cite{katz_seeding_2015} as the black diamonds, the fit of \cite{katz_seeding_2015} as the dashed line, and the functions of \cite{gieles_concurrent_2018} for two normalizations in hollow orange and pink shapes. The orange (N1) shows a reference VMS mass of $10 M_\odot$ and $N= 1e5$, and the pink (N2) show a reference VMS mass of $10^{4.3}M_\odot$ and $N= 1e7$, both drawn from realizations in their study, with circles (squares, triangles) denoting the index $\delta=0.1 (0.5,0.9)$. {\bf Right:} Black hole - stellar mass relation assuming 100\% of VMS mass is converted to a black hole at the end of its life, compared to observational relation and theoretical models. The minimum black hole mass shown is $200M_\odot$. Pink circles show dwarf AGN from \cite{greene_2020_intermediate} and down arrows show dynamical upper limits from the same work. Purple squares show JWST high-redshift AGN from \cite{harikane_jwstnirspec_2023} and \cite{maiolino_jades_2024}. The yellow star denotes the Milky Way galactic center \citep{genzel_galactic_2010}. The orange circles show local AGN from \cite{reines_relations_2015}. The blue stars show possible IMBHs, including $\omega$ Centarui \citep[][]{haberle_fast-moving_2024}, NGC 205 \citep[][]{nguyen_improved_2019}, NGC 4395 \citep[][]{woo_10000-solar-mass_2019}, G1 \citep[][]{gebhardt_20000_2002} and B023-G078 \citep[][]{pechetti_detection_2022}. The dark grey squares show the $M_{\rm BH}-M_{\rm *, galaxy}$ relation of \cite{kritos_supermassive_2024}, while the lighter grey circles show the $M_{\rm BH}-M_{\rm *, NSC}$ relation of \cite{kritos_supermassive_2024}. The grey lines show $M_{\rm BH}=0.1M_*, 0.01M_*,0.001M_*,0.0001M_*$ from top to bottom. }
    \label{fig:vmsmass}
\end{figure*}
This  analysis agrees the findings of many previous studies \citep[e.g.,][]{Sakurai17,reinoso_formation_2023,vergara_efficient_2025,pacucci_little_2025,vergara_rapid_2025, gonzalez_prieto_intermediate-mass_2024, mestichelli_teen_2026}--namely, that dense systems can host runaway merger processes which may form a very massive star.

Using the method outlined in \S \ref{sec:VMS_formation}, we estimate the resulting final VMS mass for our wide parameter space of systems. 
We show systems $M_* = 1M_\odot$ and $\alpha = 1.2$ run to $z=5$ for the range of radius and mass in \autoref{tab:parameters}, using our defaults of the strict merger criterion and $\phi_{\rm bin}=0.2, e=0.5$. 
In the left panel of \autoref{fig:vmsmass}, we show the results as a function of density. 
In some cases, less than $10M_\odot$ of collision products end up the  ``VMS," but the final mass strongly depends on density, in some cases reaching over $10^5M_\odot$. 
At fixed density, the final VMS mass is a function of the star cluster mass (shown in the color bar). 

The final VMS mass is most sensitive to density, but also depends on other parameters. For example, when we run the same systems with $M_* = M_{\rm coll} = 17M_\odot$, representing a top heavy IMF, only 22 systems form a VMS, and the most massive is $512M_\odot$. This is because the dominant disruption time is the short main sequence lifetime of the massive stars.
Meanwhile, if we increase $\alpha$ from $1.2$ to $1.75$, the maximum VMS mass increases by an order of magnitude due to the increased central concentration of stars.

For fixed $\alpha=1.2$ and typical stellar mass of $1M_\odot$, we find that the threshold to form a VMS greater than $200M_\odot$ is $\rho_h \gtrsim10^5M_\odot$ pc$^{-3}$. Above this threshold, nearly all our lowest mass systems are able to form a VMS.
Below the threshold, VMS formation is still possible for massive clusters with $M\gtrsim10^6M_\odot$, as high density and rapid collisions in their core may still allow for the runaway process to occur, but this is sensitive to the chosen parameters. 

In the right panel, we plot the $M_{\rm BH}-M_*$ relation, assuming that any VMS objects with mass above $\sim200M_\odot$ are converted into a black hole of equal mass (i.e., 100$\%$ of mass fraction retained). 
As implied by the left panel, the ratio between the black hole mass and the stellar mass depends on the system's density. 
If systems reach extremely high density, the initial black hole is already overmassive compared to the local relation. 
However, lower density systems lie comfortably along the local relation.

\subsection{Systems with black holes}
\label{sec:black holes}
\begin{figure*}
    \centering
    \includegraphics[width=0.8\linewidth]{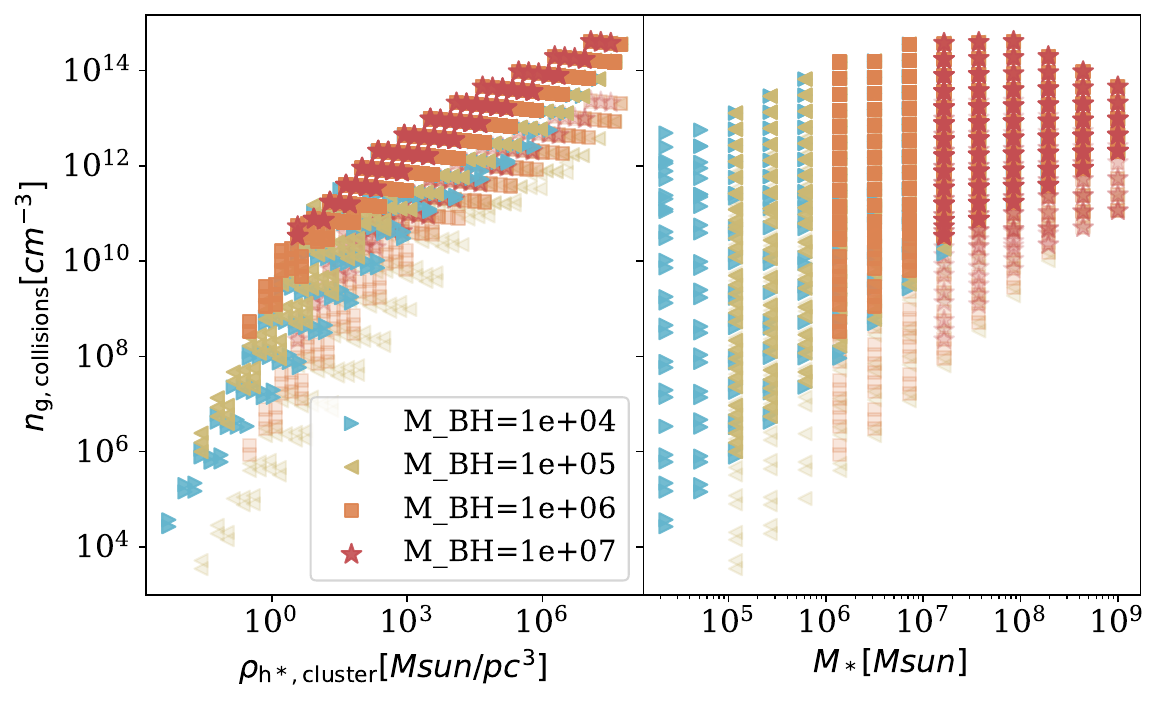}

    \caption{Average density of gas within $10^{-5}$pc surrounding the massive black hole produced by destructive collisions versus  cluster half-mass stellar density (left) and stellar mass (right). Four black hole masses are plotted: $10^4 M_\odot$ (blue rightward triangles),  $10^5 M_\odot$ (beige leftward triangles),  $10^6 M_\odot$ (orange squares), and $10^7 M_\odot$ (red stars). The average density within $10^{-4}$pc is shown in in the transparent points. }
    \label{fig:rhobhs}
\end{figure*}

For star clusters that host a black hole, our timescale analysis suggests that collisions do occur within the main sequence lifetime of stars (see black hole panel in \autoref{fig:exampleNcoll}). 
Unlike the stellar-only case, the black hole dominates the potential in the inner regions and causes collisions to be destructive. 
For most systems, we find that there are also significant constructive collisions in the outskirts of the cluster. 
For the purposes of this work, we consider the impact of destructive collisions on the environment of the black hole.

From \S \ref{sec:model_bh}, we calculate the total mass of gas from stellar interiors that is injected into the region by destructive collisions. 
We plot the average number density of gas produced by destructive collisions within $10^{-5}$pc and $10^{-4}$pc in \autoref{fig:rhobhs}.
From the plot, it is clear that systems will accumulate extremely high density gas in the innermost regions through collisions.
When using a $17M_\odot$ population representing a top heavy IMF, we find that the density in the inner $10^{-5}$pc is reduced by roughly one order of magnitude, due to the short lifetime of these stars.
The fact that the densities are still high even with the extremely short stellar lifetime highlights the rapid pace at which destructive collisions accumulate at high redshift.

\section{Discussion}
\label{sec:comparisontheoryandobs}
\subsection{Comparison to observed high redshift systems}

\begin{figure*}
    \centering
    \includegraphics[width=0.75\linewidth]{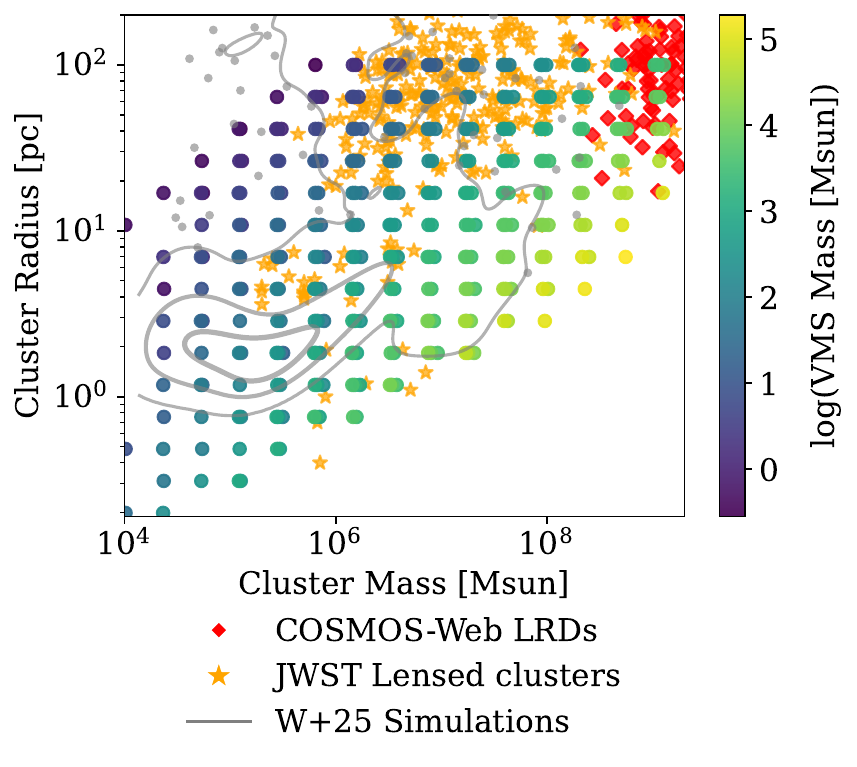}
    \caption{Mass versus radius, colored by VMS mass, with simulated systems from the cosmological simulations of \cite{Williams+25} underlaid beneath the analytic points in grey contours. The red points show Little Red Dots from \cite{akins_2025_cosmosweb}  and the orange stars show lensed clusters observed with JWST, taken from \cite{Adamo+24, claeyssens_star_2023,claeyssens_tracing_2025, fujimoto_primordial_2024, messa_anatomy_2025, Mowla+24, Vanzella+23}. The Figure highlights that observations are broadly consistent with a highly collisional scenario in the early Universe. }
    \label{fig:comparesystems}
\end{figure*}
In \autoref{fig:comparesystems}, we compare the mass and radius of the star clusters in our model (i.e., the left hand track of the flowchart in \autoref{fig:flowchart}) to a variety of observed and simulated systems.
The red diamonds represent the catalog of LRDs from \cite{akins_2025_cosmosweb}.
The orange stars are a variety of lensed star clusters observed with JWST at $z>5$. 

Given the models shown in \autoref{fig:comparesystems}, we see that the implied densities of the LRD systems would lead to the formation of very massive stars in the range of $10^4-10^5M_\odot$, in agreement with \cite{pacucci_little_2025}. 
Thus, we concur with \cite{pacucci_little_2025} that dynamically speaking, these environments are highly likely to host black holes. 
Recent studies of LRD systems suggest that the observational properties of these systems may be explained by an actively accreting black hole embedded in dense gas on nuclear scales \citep[e.g.,][]{naidu_black_2025, maiolino_jwst_2025, kido_black_2025, rusakov_little_2026, sneppen_inside_2026}.
These models include a ``black hole star," with a dense, turbulent stellar-like atmosphere surrounding the central SMBH \citep[e.g.,][]{naidu_black_2025}, or a black hole surrounded by a compact, massive envelope of optically thick gas \citep[e.g.,][]{kido_black_2025}. 
Alternatively, a less massive ``cocoon" of gas may reprocess the emission primarily through scattering 
\citep[e.g.,][]{rusakov_little_2026, sneppen_inside_2026}. 
We note that if a black hole forms or resides inside one of these stellar systems, stellar collisions will rapidly provide a supply of dense gas to the innermost regions.
Furthermore, \cite{rusakov_little_2026} note that LRD lines may be broadened by scattering with a narrow intrinsic core corresponding to $10^5-10^7M_\odot$, overlapping with the predicted masses of BHs formed due to collapse of a VMS in \autoref{fig:vmsmass}. 
Thus, dense clusters may provide the environment to not only build a black hole through rapid collisions but also generate a ready supply of dense gas to feed to the nuclear region.

The grey contours represent $z=12$ star clusters and galaxies from the AREPO cosmological simulations of \cite{Williams+25}.
Both the observational and simulated catalogs are limited in mass--for the cosmological simulations, this lower limit is just above $10^4M_\odot$. 
Additionally, the cosmological simulations are limited to $r>1$pc due to resolution. 
Even with these caveats, the parameter space of simulated models includes a significant number of clusters where VMS formation occurs. 
As explored in \cite{Williams+25, williams_observing_2026}, high density environments form naturally at high redshift within $\Lambda$CDM. 
Comparing to observed star clusters--shown as yellow stars--we see that we may have already observed environments on the small scale where collisions may have been present. 
Some of the lensed clusters may host or have hosted a VMS of $10^3-10^4M_\odot$, depending on the intrinsic density profile (see also recent work by \citealt{schleicher_massive_2026}).

\cite{bellovary_little_2025} suggest that LRDs are ``factories" of TDEs, producing roughly $10^{-4}$ events per year to explain the observed population of LRDs. 
\cite{kritos_supermassive_2024, kritos_nuclear_2025} similarly frame LRDs in terms of a nuclear star cluster model, and predict that TDEs will occur at 10x the rate of Extreme Mass Ratio Inspirals in LRDs. 
While we leave an estimate of the TDE rate to future work, we note that in our systems containing a black hole, there is a significant region of constructive collisions with fast dynamical friction outside the mass loss radius, and therefore systems excited to low angular momentum orbits, which would indeed produce TDEs \citep[e.g.,][]{rose_orbital_2025}. 

\subsection{Nitrogen enrichment}

In the JWST era, observations of lensed star clusters are providing an unprecedented view into the formation of globular clusters \citep[e.g.,][]{Adamo+24, Mowla+24,messa_anatomy_2025}.
It seems increasingly likely that the environments of globular cluster formation are being observed at high redshift, as enriched Nitrogen in accordance with local G2 populations is seen in high-$z$ clusters \citep[e.g.,][]{pascale_is_2025}. 
Many objects that naturally form in simulations--if taken as initial conditions for the analytic model here--will accumulate a significant reservoir of gas within a short period of time with elevated CNO ratios. 
The relationship between runaway stellar collisions and globular cluster nitrogen enrichment has been investigated in detail in previous works, including \cite{gieles_concurrent_2018}, \cite{vink_very_2023} and \cite{charbonnel_n-enhancement_2023}.
As in these works, we show that VMS formation may lead to injection of unusually nitrogen enriched gas into the cluster.  
Coupling the analytic models explored here to the ``initial conditions" provided by simulations, we additional suggest that significant stellar material released through constructive collisions will pile up (even in clusters where the typical collision is not destructive).
This gas may reside in the star cluster region unless feedback can expel it, resulting a physical picture similar to the model of \cite{pascale_nitrogen-enriched_2023}.

\subsection{Limitations of this model}
Here, we briefly discuss the important limitations of the model presented in this study. 
In this work, we analyze a broad range of systems analogous to high-redshift observations and simulations, aiming to find the regions of parameter space where collisions may lead to dynamically interesting outcomes, especially the formation of a VMS or Little Red Dot analogue. With this analytical approach, we do not reproduce the exact scenario of any given system.
In particular, our method of estimation of the mass loss fraction, the neglecting of gas in the star clusters, and the termination of the systems upon merger may all lead to the underestimation of dynamical friction and collision rates. 
These effects may be counteracted by the fact that for the no-black-hole systems, we do not capture the restructuring of the density profile due to the short relaxation times. 
This may lead to an overestimation of the dynamical friction and collision rates in those cases.  
In the paragraphs below, we discuss these effects in detail and compare our methodology and results to other works in the literature. 

\paragraph{The mass loss fraction:} 
Our estimate of the fractional mass loss over the course of a collision is based on the energy ratio between the collision and the stars' binding energies \citep[following][]{lai_collisions_1993, rose_stellar_2023}.
More sophisticated approaches take into account the impact parameters and detailed stellar structure involved in the collisions. 
Of particular concern would be the transition radius from constructive to destructive collisions. Here, we compare the stars' dispersion velocity and their individual escape velocity. Common mass loss/gain prescriptions often depend on the collision velocity  \citep[e.g.,][]{lai_collisions_1993, rauch_collisional_1999, rose_stellar_2023, rose_modeling_2026}.
\cite{rose_stellar_2023} compared models of the mass loss fraction in the literature using the Monte Carlo method described here, and found that the region in which destructive collisions dominated was roughly consistent across various prescriptions, coinciding with the transition radius given by equating the velocity dispersion to the escape speed from the star (\autoref{eq:destructivecriterion}).  
%
For the value of the mass loss fraction, we can compare to more detailed simulations of this effect. 
For instance, compared to \cite{rose_modeling_2026}, we see that typical mass loss fractions are smaller than the estimate given here. 
Thus, we likely overestimate the destructiveness of collisions and the resultant quantity of gas in these environments. 
For systems with a black hole, this may imply that we overestimate the densities achieved by gas originating in destructive collisions. 
However, balancing this overestimation of gas densities is the existence of an additional source of infalling  material onto the central region from outside the transition radius. 
This is provided by constructive collisions.
%
Thus, we reserve a more complex treatment of collision outcomes for future work.

\paragraph{Mass spectrum evolution} Our calculations assume a static stellar mass distribution, and the mass spectrum of stars is not updated to include the outcomes of constructive collisions.
This means we miss runaway growth effects, which can occur in extreme scenarios, where stellar growth enhances gravitational focusing and further collisions. 
We have chosen to remain conservative in this respect.
For the growth of a VMS in self-gravitating systems, we note that because the dynamical friction time is short, stars should move rapidly to the central regions once they experience a small number of collisions (this is reflected in our Monte Carlo simulations as well). 
For black hole hosting systems, we have only considered destructive collisions, and terminated further collisions at a given radius once 50\% of stars have experienced collisions (i.e., the number density of stars has changed by an order of magnitude of itself). The constructive regime for black holes, where evolving the mass spectrum would be particularly relevant, is not considered in this work as contributing to the gas density in the nuclear region. 

\paragraph{Secular approximations and comparison to previous works:}
In using the collision rate implied by \autoref{eq:t_coll}, we approximate stellar motion through a sea of stars with density described by a radial power law profile. 
In a more complete physical picture, collisions should alter this density profile, and while we include a ``depletion" term in our VMS growth estimate, the relaxation times are short enough that restructuring of the system may occur. 
Thus, we do not truly account for a full runaway collision scenario in either our analytic or Monte Carlo simulations. 
We have approached this drawback by cutting off our systems at the core collapse timescale, if that is the limiting timescale, but this physics is more accurately modeled by a full N-body approach. 
We can thus additionally compare our results to theoretical studies in the literature representing a variety of methodologies to account for evolving system's potential and structure. 
This is shown in in \autoref{fig:vmsmass}. 
\cite{gonzalez_prieto_intermediate-mass_2024} model isolated, virialized star clusters using the Cluster Monte Carlo code, with no pre-existing black hole and no external disruption timescale. 
Our model differs mostly methodologically from \cite{gonzalez_prieto_intermediate-mass_2024}, and therefore it is encouraging that despite the different approaches, the predicted VMS masses are generally consistent at fixed cluster mass and density for low densities. 
At high densities, our simulations of comparable mass lie above their best-fit curve.
This is likely due to the additional concentration of mass in the center of the power-law profile in this work compared to the Plummer profile used in the initial conditions of \cite{gonzalez_prieto_intermediate-mass_2024}. 
The TITANS suite \citep{mestichelli_teen_2026} use direct N-body simulations of clusters with masses in the range of $10^5-10^6M_\odot$ with properties similar to low-redshift YMCs. 
Their VMS masses form typically through primordial binary mergers rather than collision chains, which are absent in their simulation when $\rho_h<500M_\odot$ pc$^{-3}$. 
We note that below this threshold in density, no systems with comparable mass to the \cite{mestichelli_teen_2026} simulations form a star above 100$M_\odot$, consistent with their simulations.
\cite{pacucci_little_2025} focus on systems with more extreme central densities ($10^4-10^9M_\odot$ pc$^{-3}$). 
Their analytic model, Fokker Planck solver, and direct N-body methods generate VMS with $M\sim 9\times10^3-5\times10^4$, which coincide with our predictions for the highest density systems we consider. 
The consistency of their three methods and our independent set of models suggests that the equilibrium VMS mass is robust in this extreme regime. 
Similarly, \cite{vergara_rapid_2025} simulates a single ultra-compact cluster with full stellar evolution. 
\cite{vergara_rapid_2025} resolve individual collision chains, and the resulting VMS mass for that realization agrees with our integrated mass estimate. 

Several analytical studies model the problem of black hole seeding and VMS growth \citep[e.g.,][]{devecchi_formation_2009, katz_seeding_2015, gieles_concurrent_2018}. 
\cite{devecchi_formation_2009} and \cite{katz_seeding_2015} do not include radial profile integration or gravitational focusing with eccentricity, effects which will increase the projected rate of runaway collisions, or wind-limited VMS loss. 
Our models at fixed density tend to have increased VMS mass in \autoref{fig:vmsmass}, consistent with this additional physics. 
\cite{gieles_concurrent_2018} model globular cluster formation in converging gas flows where gas accretion onto protostars contributes to increased stellar radii and contraction of the cluster, increasing the collision cross section (as we describe below). 
From the Figure, \cite{gieles_concurrent_2018} predict that the VMS mass will decrease with increasing density. 
Importantly, the density of the cluster plotted is the peak density reached during contraction (an output, rather than an input of their model).
For their systems, a higher peak density leads to rapid two-body heating which terminates collisions in their clusters.
However, we include disruption times that in some cases do not depend on density (and additionally do not model gas accretion and core contraction/heating). 
Like \cite{gieles_concurrent_2018}, \cite{boekholt_formation_2018} model the protostellar regime where gas accretion inflates stellar radii and enhances collision rates using an N-body simulation with the `sticky spheres' approximation. 
They achieve final masses in the range of $10^4-10^5M_\odot,$ for systems of initial mass between $10^4-10^6M_\odot,$ consistent with our most extreme systems. 
This suggests that our gas-free model should potentially be treated as a lower limit for what likely are gas-rich environments. 
Recently, \cite{schleicher_massive_2026} also suggested that observed dense clusters may host IMBH candidates, considering a single-zone timescale comparison applied directly to observed cluster properties (rather than the radially resolved integral considered here). 
They suggest that 16\% of observed young massive clusters may runaway collisions leading to IMBHs. 
Comparing to our \autoref{fig:comparesystems}, where we predict the achieved mass of VMS that may evolve to an IMBH, we find that most observed clusters (orange stars) form only modest VMSs, with only the most compact systems achieving $M_{\rm VMS}\gtrsim 10^3-10^4M_\odot$.
This is broadly consistent with \cite{schleicher_massive_2026}.

The previous studies do not include possible pre-existing IMBHs or SMBHs in the stellar systems. \cite{kritos_supermassive_2024} embed nuclear star cluster (NSC)-based BH seeding in a full cosmological merger tree (whereas we are agnostic as to whether the clusters hosting black holes are nuclear star clusters). 
They use the \cite{zwart_runaway_2002} single-zone criterion for a binary condition on whether collisional runaway and therefore seed formation occurs. Meanwhile, we compute the VMS mass continuously as a function of the cluster parameters. The offset between our points and their NSC relation in the right panel of \autoref{fig:vmsmass} is natural due to their inclusion of the continued growth of the SMBH. As discussed, we find that the resulting black holes will be embedded in dense gas, and will likely continue to be bombarded with TDEs--thus, we expect that they should experience further growth.
However, we do not model the subsequent accretion onto the black hole. 

\paragraph{Effects of gas:} As discussed above, we do not take into account the effects of gas in the region of the star cluster. 
From the initial set of simulations in \cite{Williams+25}, we cannot get a good estimate of the initial gas mass since stellar feedback is not included. 
However, it is clear that gas leftover from the formation of the star clusters may indeed be present in the region \citep[e.g.,][]{chen_supersonic_2025,Lake+24a, Williams+23}.
Furthermore, recent work has shown the importance of even a small fraction of gas to the dynamical evolution of a system \citep[e.g.,][]{rozner_stellar_2025, rozner_binary_2023}. 
Since the effect of gas will be to steepen the stellar density profile, \citep[][]{rozner_stellar_2025}, we conclude that the impact of neglecting gas effects is to underestimate the rate of dynamical friction driving migration into the inner regions. 
\cite{schleicher_massive_2026} suggest that above $\sim6\times 10^6M_\odot$, clusters can retain gas in spite of supernova feedback and grow the central object through gas dynamical friction and Bondi accretion.
Thus, we suspect that the inclusion of gas would increase the accretion rates onto the growing VMS in this model.

\paragraph{Mergers with the cosmological environment:}
In our model, a merger with a nearby halo is considered a "disruptive" process, and we terminate the secular timescale arguments at that time. 
However, it is possible that mergers could drive important behaviors that increase the rates of the dynamical processes considered here through violent relaxation or injection of gas into the region. 
On the galaxy scale, mergers can produce the formation of additional dense clusters \citep[e.g.,][]{nakazato_merger-driven_2024}. 
We leave the investigation of such situations to future studies. 

\section{Summary}
\label{sec:summary}
\begin{enumerate}
    \item We introduce a radially resolved analytic model for stellar collisions in high redshift systems with or without central black holes. 
    The model is shown schematically in \autoref{fig:schematic} and \autoref{fig:flowchart}. We account for destructive or constructive collisions, and estimate the mass of VMS that may form in the cluster cores due to dynamical migration of massive merger products. We also estimate the buildup of dense gas in the nuclear region of black holes due to destructive collisions. We use our model to consider a wide range of star cluster parameters, including size, mass, density profile, BH mass, eccentricity, binary fraction, virialization state, and IMF. 
    \item Even with our strictest combination of parameters, stellar collisions should be dynamically important in many compact systems at high redshift ($12\gtrsim z \gtrsim5$). 
    Observations and cosmological simulations suggest that such environments are indeed present at high redshift (see \autoref{fig:comparesystems}). Many systems experience collisions in some fraction of their central region, and a subset of these experience collisions within the mass segregation regime, suggesting that collision products will sink to the center (see for reference \autoref{fig:exampleNcoll}). 
    \item Rapid formation of a very massive star (VMS) will occur in dense systems without a massive black hole where collisions are typically constructive (as depicted in Figure \autoref{fig:vmsmass}). These VMSs may evolve into intermediate mass and supermassive black holes.
    \item Systems hosting a supermassive black hole will typically experience destructive collisions in the nuclear regions, leading to the buildup of dense gas in the vicinity of the black hole on rapid timescales (e.g., \autoref{fig:rhobhs}). 
    Systems where collisions are common will likely be somewhat gas enriched in their central regions, whether collisions were destructive or constructive. This gas--originating in the interiors of main sequence stars--may display unusual metal enrichment. 
    \item Population III star clusters--if top heavy--should not typically experience significant constructive stellar collisions on the main sequence lifetime. However, they can still produce dense gas in the nuclear region through destructive collisions if a massive black hole is present in these star clusters.
\end{enumerate}

\section*{Acknowledgments}

C.E.W. acknowledges the support of the National Science Foundation Graduate Research Fellowship, the University of California, Los Angeles (UCLA), the Mani L. Bhaumik Institute for Theoretical Physics, and the UCLA Center for Developing Leadership in Science Fellowship. C.E.W., W.L., S.N., B.B., F.M., and M.V. thank the support of NASA grant No. 80NSSC24K0773 (ATP-23-ATP23-0149) and the XSEDE/ACCESS AST180056 allocation, as well as the UCLA cluster Hoffman2 for computational resources. C.E.W and S.N. thank Howard and Astrid Preston for their generous support. B.B. also thanks the Alfred P. Sloan Foundation and the Packard Foundation for support.
N.Y. and C.E.W. acknowledge financial support from JSPS International Leading Research 23K20035.
 F.M. acknowledges support from the
European Union—NextGeneration EU within PRIN 2022
project n.20229YBSAN—Globular clusters in cosmological
simulations and in lensed fields: from their birth to the present
epoch.
This material is based upon work supported by the National Science Foundation Graduate Research Fellowship Program under Grant No. DGE-2034835. Any opinions, findings, conclusions, or recommendations expressed in this material are those of the author(s) and do not necessarily reflect the views of the National Science Foundation.
This work used computational and storage services associated with the Hoffman2 Cluster, which is operated by the UCLA Office of Advanced Research Computing’s Research Technology Group.

\software{astropy \citep{2013A&A...558A..33A,2018AJ....156..123A}, matplotlib \citep{Matplotlib},  numpy \citep{harris2020numpy}, scipy \citep{2020SciPy-NMeth}.}

\bibliography{cosmo}{}
\bibliographystyle{aasjournalv7}

\appendix
\section{Exact solutions}
\label{sec:exactsolutions}
\subsection{Number of collisions - full solutions}

Assuming a power law density profile with index $\alpha$, \autoref{eq:GeneralNCollisions} evaluates to the following expression:  
\begin{multline}
    N_{\rm coll, tot} =  \frac{\pi t_d f_{M_*}^{\rm IMF}(3-\alpha)}{M_*^2} \bigg[ 
    F_1(e) \sqrt{\frac{c_v G}{1+\alpha}}\frac{(4\pi)^{3/2}\rho_0^{5/2}}{r_0^{-5\alpha/2}(3-\alpha)^{3/2}}
    \left(\frac{1}{(4-5\alpha/2)} r^{4-5\alpha/2}\right)\\
    +F_2(e)\left(\frac{c_v G}{1+\alpha}\right)^{-1/2}
    \frac{(4\pi)^{1/2}\rho_0^{3/2}}{r_0^{-3\alpha/2}(3-\alpha)^{1/2}} 
    \left(\frac{1}{(2-3\alpha/2)}r^{2-3\alpha/2}\right)\bigg]\bigg|_{r_{\rm min}}^{r_{\rm max}}
    \label{eq:plNumberofCollisions}
\end{multline}
This equation only converges at $r=0$ if $-1<\alpha\lesssim1.333$. For $\alpha =1.6$ and $\alpha = 4/3$, the expression is discontinuous, so we thus re-integrate for these special cases and provide the expressions in \autoref{sec:exactsolutions}.

For the addition of a black hole, we use the velocity dispersion given by \autoref{eq:sigma_bhpl}, and solve the expression in \autoref{eq:GeneralNCollisions}, giving:
\begin{multline}
      N_{\rm coll, tot} =\frac{\pi t_d f^{\rm IMF}_{M_*}(3-\alpha)}{M_*^2}\Bigg[F_1(e) \left(\frac{c_v G}{(1+\alpha)} \right)^{1/2}  c_\rho c_M\frac{\sqrt{M_{\rm BH}}r^{5/2-2\alpha} {}_2F_1(\frac{1}{2};\frac{5/2-2\alpha}{3-\alpha};\frac{11/2-3\alpha}{3-\alpha};-\frac{c_Mr^{3-\alpha}}{M_{\rm BH}}) }{5/2-2\alpha}\\
    +F_2(e) \left(\frac{c_v G}{(1+\alpha)} \right)^{-1/2} c_\rho c_M 
 \frac{\sqrt{\frac{1}{M_{\rm BH}}}r^{7/2-2\alpha} {}_2F_1(\frac{1}{2};\frac{7/2-2\alpha}{3-\alpha};\frac{13/2-3\alpha}{3-\alpha};-\frac{c_Mr^{3-\alpha}}{M_{\rm BH}}) }{7/2-2\alpha}\Bigg]\Bigg|_{r_{\rm min}}^{r_{\rm max}}
    ,
    \label{eq:BHncolltot}
\end{multline}
where
\begin{align}
    c_M & = \frac{4\pi \rho_0}{(3-\alpha)r_0^{-\alpha}}, \\
    c_\rho & = \frac{\rho_0}{r_0^{-\alpha}}.
\end{align}
We note that this expression is discontinuous in the important cases of $\alpha = 5/4$ and $\alpha = 7/4$. 
We find exact expressions for these cases, which are provided in below.

\subsubsection{Discontinuous cases}
\paragraph{$\alpha=1.25$, power law with black hole}

When $\alpha =1.25$, it is clear that the first hypergeometric term of \autoref{eq:BHncolltot} has a singularity in the denominator. Retaining the non-discontinuous solution for the second term, we integrate \autoref{eq:GeneralNCollisions} specifically for this case to avoid this discontinuity, finding: 

\begin{multline}
       N_{\rm coll, tot}(\alpha=5/4)=\frac{\pi t_d f^{\rm IMF}_{M_*}(3-\alpha)}{M_*^2}\bigg[F_1(e) \left(\frac{c_v G}{(1+\alpha)} \right)^{1/2} c_\rho c_M \\\times
\left[
\frac{8}{7}\sqrt{c_Mr^{7/4}+M_{\rm BH}}
-\frac{8}{7}\sqrt{M_{\rm BH}}\,\sinh^{-1}\!\left(
\frac{\sqrt{M_{\rm BH}}}{\sqrt{c_M}r^{7/8}}\right)
\right]\\      
    +F_2(e) \left(\frac{c_v G}{(1+\alpha)} \right)^{-1/2}c_\rho c_M 
\frac{\sqrt{\frac{1}{M_{\rm BH}}}r^{7/2-2\alpha} {}_2F_1(\frac{1}{2};\frac{7/2-2\alpha}{3-\alpha};\frac{13/2-3\alpha}{3-\alpha};-\frac{c_Mr^{3-\alpha}}{M_{\rm BH}}) }{7/2-2\alpha}
    \bigg]\bigg|_{r_{\rm min}}^{r_{\rm max}}.
\end{multline}

\paragraph{$\alpha = 1.75$, power law with black hole}
When $\alpha = 1.75$, the second hypergeometric term of \autoref{eq:BHncolltot} is discontinuous. 
We reintegrate for this specific case, preserving the non-discontinuous first term and finding: 

\begin{multline}
       N_{\rm coll, tot}(\alpha=7/4)=\frac{\pi t_d f^{\rm IMF}_{M_*}(3-\alpha)}{M_*^2}\bigg[F_1(e) \left(\frac{c_v G}{(1+\alpha)} \right)^{1/2} c_\rho c_M\frac{\sqrt{M_{\rm BH}}r^{5/2-2\alpha} {}_2F_1(\frac{1}{2};\frac{5/2-2\alpha}{3-\alpha};\frac{11/2-3\alpha}{3-\alpha};-\frac{c_Mr^{3-\alpha}}{M_{\rm BH}}) }{5/2-2\alpha}\\
    +F_2(e) \left(\frac{c_v G}{(1+\alpha)} \right)^{-1/2} c_\rho c_M
\left[
-\frac{8}{5\sqrt{M_{\rm BH}}}\,
\sinh^{-1}\!\left(
\frac{\sqrt{M_{\rm BH}}}{\sqrt{c_M}r^{5/8}}
\right)
\right]
    \bigg]\bigg|_{r_{\rm min}}^{r_{\rm max}}
\end{multline}

\paragraph{$\alpha =1.6,$ power law}
\begin{multline}
    N_{\rm coll, tot}(\alpha=8/5)=  \frac{\pi t_d f_{M_*}^{\rm IMF}(3-\alpha)}{M_*^2} \bigg[ 
    F_1(e) \sqrt{\frac{c_v G}{1+\alpha}}\frac{(4\pi)^{3/2}\rho_0^{5/2}}{r_0^{-5\alpha/2}(3-\alpha)^{3/2}}
    \ln(r)\bigg|_{r_{\rm min}}^{r_{\rm max}}\\
    +F_2(e)\left(\frac{c_v G}{1+\alpha}\right)^{-1/2}
    \frac{(4\pi)^{1/2}\rho_0^{3/2}}{r_0^{-3\alpha/2}(3-\alpha)^{1/2}} 
    \left(\frac{1}{(2-3\alpha/2)}r^{2-3\alpha/2}\right)\bigg|_{r_{\rm min}}^{r_{\rm max}}\bigg]
\end{multline}
\paragraph{$\alpha =1.33,$ power law}
\begin{multline}
    N_{\rm coll, tot}(\alpha=4/3)=  \frac{\pi t_d f_{M_*}^{\rm IMF}(3-\alpha)}{M_*^2} \bigg[ 
    F_1(e) \sqrt{\frac{c_v G}{1+\alpha}}\frac{(4\pi)^{3/2}\rho_0^{5/2}}{r_0^{-5\alpha/2}(3-\alpha)^{3/2}}
    \left(\frac{1}{(4-5\alpha/2)} r^{4-5\alpha/2}\right)\bigg|_{r_{\rm min}}^{r_{\rm max}}\\
    +F_2(e)\left(\frac{c_v G}{1+\alpha}\right)^{-1/2}
    \frac{(4\pi)^{1/2}\rho_0^{3/2}}{r_0^{-3\alpha/2}(3-\alpha)^{1/2}} 
    \ln(r)\bigg|_{r_{\rm min}}^{r_{\rm max}}\bigg].
\end{multline}
We use these expressions for all analytic calculations when $\alpha$ is one of these discontinuous values.

\subsection{Mass inflow and outflow rates}
\label{sec:vmsfullexpression}
Including gravitational focusing, radius dependent mass loss, and radial dependence of the dynamical friction timescale, we have:
\begin{equation}
    \dot{M}_{\rm df} = \int_{r_{\rm min}}^{r_{\rm df}} \dd r N_{\rm coll/M*}(r)\frac{\dd N_{M*}}{\dd r }(M_i+M_*)(1-f_{\rm ml}(r)) \frac{1}{t_{\rm df}(r)}
\end{equation}
\begin{equation}
    = \int_{r_{\rm min}}^{r_{\rm df}} \dd r 
     \pi t_d n(r) \sigma(r)\left( F_1(e) + F_2(e)\frac{1}{\sigma(r)^2} \right)
     \frac{f_{M_*}^{\rm IMF} \frac{\dd{M(r)}}{\dd{r}}}{M_*} (M_i+M_*)\left(1-
     \frac{\mu \sigma(r)^2}{GM_i^2/R_i+GM_*^2/R_*}\right)
     \frac{q}{ t_{\rm relax}},
\end{equation}
\begin{multline}
    \dot{M}_{\rm df} =\frac{\pi G^2 t_d f^{\rm IMF}_{M_*} (M_*+M_i)q \ln{\Lambda}}{0.34 M_*}F_1(e) \frac{(3-\alpha)(1+\alpha)}{c_vG}  c_\rho^2 \frac{1}{1-2\alpha} r^{1-2\alpha}\\
+\frac{\pi G^2 t_d f^{\rm IMF}_{M_*} (M_*+M_i)q \ln{\Lambda}}{0.34 M_*} F_2(e) \frac{(3-\alpha)(1+\alpha)^2}{c_v^2G^2}\frac{c_\rho^2}{c_M}\frac{1}{-(1+\alpha)}r^{-1-\alpha}\\
    - \frac{\pi G t_d f^{\rm IMF}_{M_*}\mu (M_*+M_i) q \ln{\Lambda}}{0.34\left(\frac{M_i}{R_i}+\frac{M_*}{R_*}\right)M_*}  F_1(e) (3-\alpha) c_\rho^2c_M\frac{1}{3-3\alpha} r^{3-3\alpha}\\
     - \frac{\pi G t_d f^{\rm IMF}_{M_*}\mu (M_*+M_i) q \ln{\Lambda}}{0.34\left(\frac{M_i}{R_i}+\frac{M_*}{R_*}\right)M_*} F_2(e)\frac{(3-\alpha)(1+\alpha)}{c_v G} c_\rho^2 \frac{1}{1-2\alpha} r^{1-2\alpha}\bigg|_{r_{\rm min}}^{r_{\rm df}},
     \label{eq:fullmassaccretionratepl}
\end{multline}
Next, we consider the depletion rate: 
\begin{equation}
    \dot{M}_{\rm dep} = \int_{\rm rmin}^{\rm rdf} \Gamma_{\rm coll}M_{\rm coll}(r) 4 \pi r^2 n(r)\dd{r}
\end{equation}
\begin{equation}
    \dot{M}_{\rm dep}= \int_{\rm rmin}^{\rm rdf} \pi n(r) \sigma(r)\left( F_1(e) + F_2(e)\frac{1}{\sigma(r)^2} \right)(M_*+M_i)\left(1-
     \frac{\mu \sigma(r)^2}{GM_i^2/R_i+GM_*^2/R_*}\right)4 \pi r^2f^{\rm IMF}_{M*}n(r)\dd{r}
\end{equation}
\begin{multline}
        =\frac{4 \pi^2 (M_*+M_i)f^{\rm IMF}_{M*}c_\rho^2}{M_*^2}\bigg[ \left(F_1(e)-F_2(e)\frac{\mu}{GM_i^2/R_i+GM_*^2/R_*}\right)\left(\frac{c_vG}{(1+\alpha)}\right)^{1/2} c_M^{1/2}\\\times \frac{1}{(-5\alpha/2+4)} r^{-5\alpha/2+4}\bigg|_{\rm rmin}^{\rm rdf}
    \\-F_1(e)\frac{\mu}{GM_i^2/R_i+GM_*^2/R_*}\left(\frac{c_vG}{(1+\alpha)}\right)^{3/2} c_M^{3/2}\frac{1}{(-7\alpha/2+6)}r^{-7\alpha/2+6} \bigg|_{\rm rmin}^{\rm rdf} 
    \\+F_2(e)\left(\frac{c_vG}{(1+\alpha)}\right)^{-1/2}c_M^{-1/2} \frac{1}{(-3\alpha/2+2)}r^{-3\alpha/2+2}\bigg|_{\rm rmin}^{\rm rdf}
    \bigg]
    \label{eq:fulldepletionrate}
\end{multline}
And finally, the binary contribution:

\begin{equation}
    \dot{M}_{\rm bin}\approx \frac{1}{\eta\sigma_{r= rdf}^2}\int_{\rm rmin}^{\rm rdf}  \frac{G^2M_*}{\sigma(r)}\rho(r)^2(\mu_{bs}+\mu_{bb} )4 \pi r^2 \dd{r}
\end{equation}

\begin{equation}
    =\frac{4 \pi G^2 M_*^2}{\eta\sigma_{\rm rdf}^2} (\mu_{bs}+\mu_{bb} ) \left(\frac{(1+\alpha)}{c_vG}\right)^{1/2}c_M^{-1/2}c_\rho^2   
    \frac{1}{(2-3\alpha/2)}r^{2-3\alpha/2} \bigg|_{\rm rmin}^{\rm rdf}
    \label{eq:fullbinaryrate}
\end{equation}

\section{Disruption of systems}

\begin{figure*}
    \centering
    \includegraphics[width=0.48\linewidth]{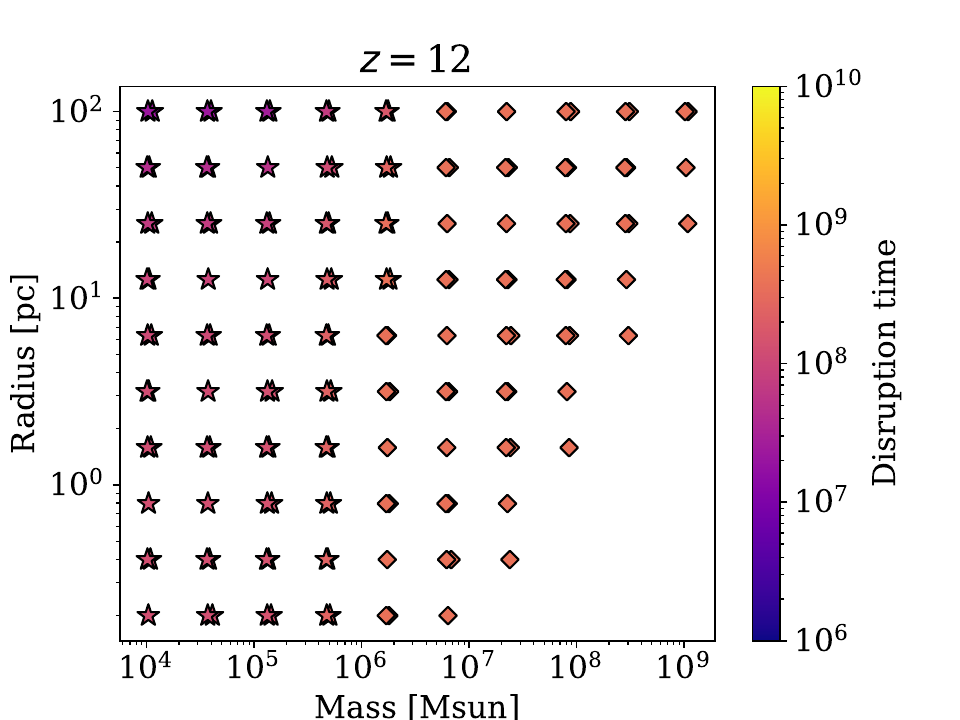}
    \includegraphics[width=0.48\linewidth]{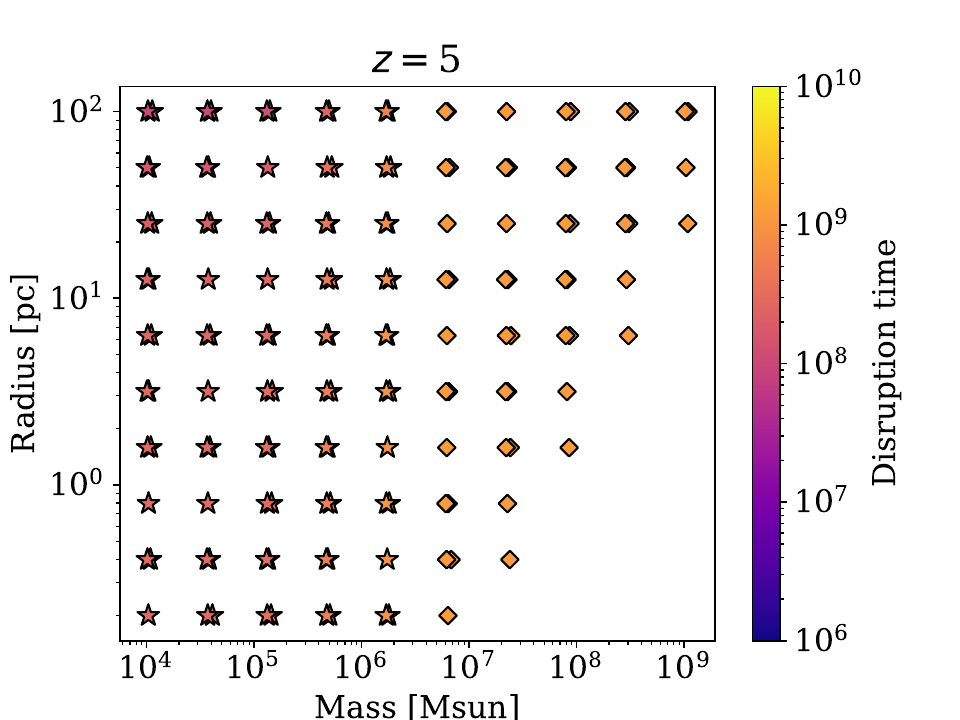}
    \includegraphics[width=0.48\linewidth]{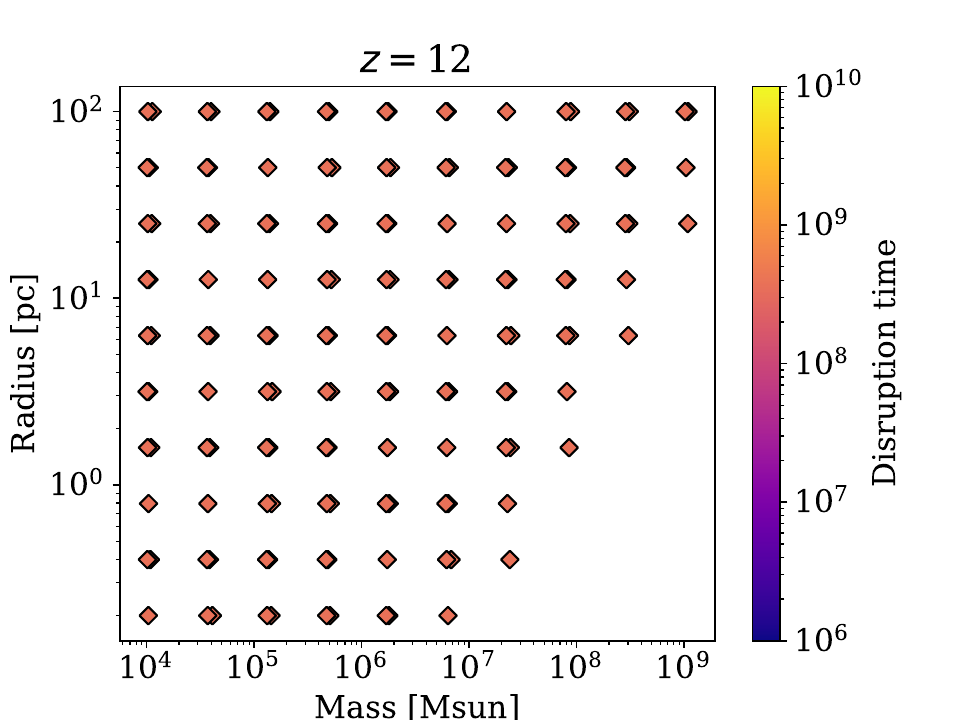}
    \includegraphics[width=0.48\linewidth]{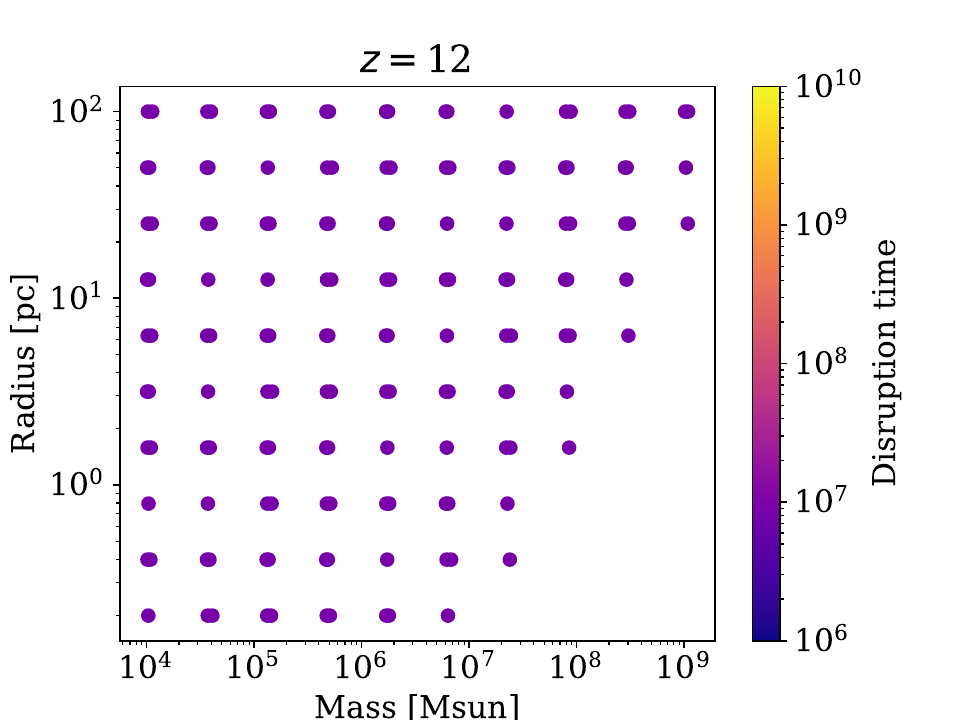}
    \caption{Disruption time (\autoref{eq:tdisrupt}) for example systems, depicted in mass-radius parameter space. Systems assume $\alpha =1.25$ with a Salpeter IMF at $z=12$ with a nearest-neighbor merger, except for: Top right panel - $z=5$; Bottom left panel - Local 10 kpc averaging used for halo merger time; Bottom left panel - top heavy IMF. Circles show where the system is disrupted by the main sequence lifetime, diamonds show systems that are disrupted by the age of the Universe, and stars show systems that are disrupted by a merger.}
    \label{fig:disruption}
\end{figure*}

In \autoref{fig:disruption}, systems (in the mass-radius parameter space) are colored by their disruption timescale (\autoref{eq:tdisrupt}). 
The marker shapes show how systems were disrupted - through mergers (stars), the main sequence lifetime running out (circles), or the age of the Universe (diamonds). 
Four scenarios are shown. 
In the top left, $z=12$ objects are assumed to have a Salpeter IMF and $\alpha=1.25$, and are disrupted by a merger with their nearest neighbor. The top right shows the same setup at $z=5$.
As is expected by our merger timescale (\autoref{eq:tmerger}) low mass and large radius systems are most likely to be disrupted by a merger. 
At high redshift, the age of the universe is relatively short compared to the main sequence lifetime at $1 M_\odot$, so high mass/ compact systems are typically "disrupted" by the limited time they've had to evolve at these early times. 
Even for those systems which are cut short by mergers, most systems have at least 10-100 Myr to evolve. 

The bottom left panel shows the effect of relaxing the merger timescale - in particular, the density in the merger timescale is now set to the average density in the region. Now, every system has the age of the Universe to evolve. 
In reality, the merger timescale varies between these extremes, with some isolated systems more accurately represented by this longer timescale. 
Thus, estimating the merger timescale using the nearest neighbor criterion will lead to an underestimate of the number of collisions. 

The most significant impact on the disruption timescale is shown in the bottom right panel of \autoref{fig:disruption}, where the cluster is assumed to have a top heavy IMF with average mass 17$M_\odot$. 
Now, every system is limited by the short main sequence lifetime of these stars, and systems have $\sim10$ Myr or less to experience collisions.

\section{Monte Carlo Simulations}
\label{app:montecarlo}
\begin{figure}
    \centering
    \includegraphics[width=0.45\linewidth]{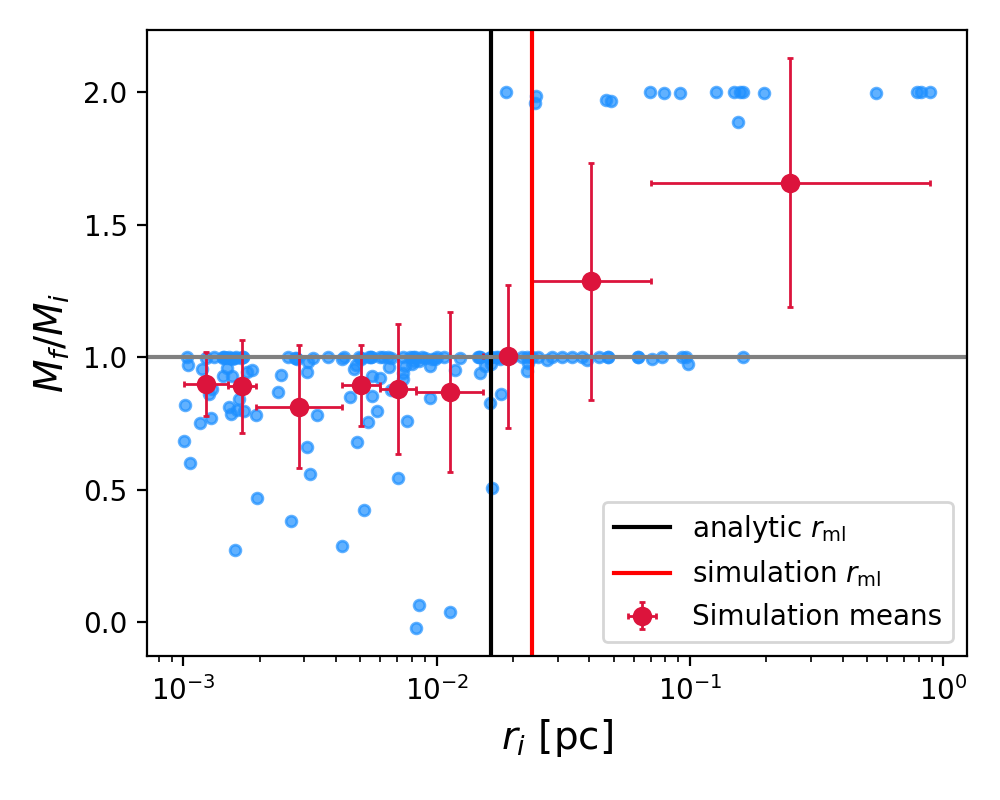}
    \includegraphics[width= 0.45\linewidth]{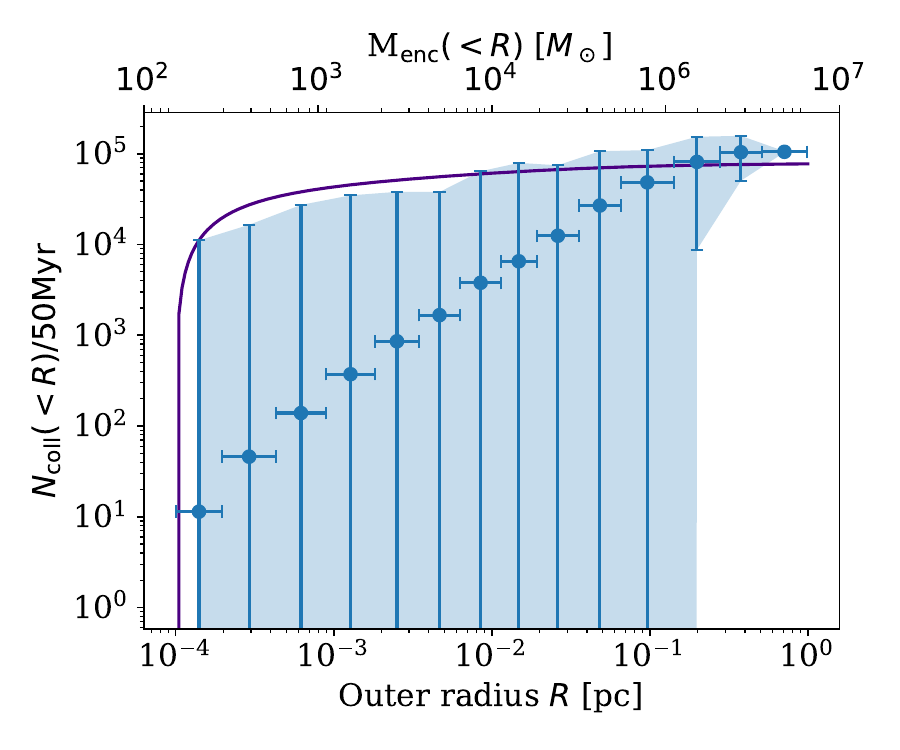}
    \caption{Left: Example realization of the Monte Carlo Simulation, showing the ratio of final to initial mass against the initial radius of the particle. Systems below the solid line lost mass, while systems above the solid line gained mass. The black line shows the analytic mass loss radius (\autoref{eq:destructivecriterion}) while the red line shows the maximum radius within which $99\%$ of systems lost mass in the course of  collisions. Right: Simulated number of collisions collisions versus analytic number of collisions versus radius for an example system containing a $10^5 M_\odot$ black hole. }
    \label{fig:simulated_collisions}
\end{figure}

\autoref{fig:simulated_collisions} shows an example system realized through the Monte Carlo numerical simulation described in \S~\ref{sec:montecarlo}. 
The left panel shows the change in stellar mass ($M_f/M_i$) versus radius, whereas the right panel shows the cumulative total number of collisions that occurred at each radius. 
We have estimated the simulation transition radius between the destructive and constructive regimes by calculating the maximum radius within which 99\% of systems have lost mass in the course of collisions, rather than gained mass. 
This radius is plotted in red. 
The analytic expression is plotted in black. 

\end{document}